\documentclass{article}
\usepackage{graphicx}
\usepackage{emulateapj,timesfonts} 

\submitted{
\it To appear in \rm The Astrophysical Journal, 509:\#\#-\#\#, 1998 December 10}

\makeatletter
\@ifundefined{chapter}{\def\thebibliography#1{
\section*{References}
  \list
  {\relax}{\setlength{\labelsep}{0em}
        \setlength{\itemindent}{-\bibhang}
        \setlength{\itemsep}{\parskip}
        \setlength{\parsep}{0pt}
        \setlength{\leftmargin}{\bibhang}}
    \def\newblock{\hskip .11em plus .33em minus .07em}
    \sloppy\clubpenalty4000\widowpenalty4000
    \sfcode`\.=1000\relax}}%
{\def\thebibliography#1{
  \list
  {\relax}{\setlength{\labelsep}{0em}
        \setlength{\itemindent}{-\bibhang}
        \setlength{\itemsep}{\parskip}
        \setlength{\parsep}{0pt}
        \setlength{\leftmargin}{\bibhang}}
    \def\newblock{\hskip .11em plus .33em minus .07em}
    \sloppy\clubpenalty4000\widowpenalty4000
    \sfcode`\.=1000\relax}}

\newlength{\bibhang}
\makeatother

\newcommand{\BC}{$B/C$}
\newcommand{\Berat}{$^{10}Be/\,^9Be$}
\newcommand{\Dpp}{D_{pp}}
\newcommand{\Dxx}{D_{xx}}
\newcommand{\ddp}{{\partial\over\partial p}}
\newcommand{\dfdt}{{\partial f(\vec p)\over\partial t}}
\newcommand{\dfdp}{{\partial f(p)\over\partial p}}
\newcommand{\dNpdt}{{\partial \psi\over\partial t}}
\newcommand{\dNpdp}{{\partial \psi\over\partial p}}

\newcommand{\zii}{z_{i}}
\newcommand{\zip}{z_{i+1}}
\newcommand{\zim}{z_{i-1}}
\newcommand{\pii}{p_{i}}
\newcommand{\pip}{p_{i+1}}
\newcommand{\pim}{p_{i-1}}

\newcommand{\Dip}{D_{pp,i+1}}
\newcommand{\Dim}{D_{pp,i-1}}
\newcommand{\Nii}{\psi_{i}}
\newcommand{\Nip}{\psi_{i+1}}
\newcommand{\Nim}{\psi_{i-1}}
\def\fwb{80mm}
\def\fwc{55mm}

\lefthead{STRONG \& MOSKALENKO}
\righthead{PROPAGATION OF COSMIC-RAY NUCLEONS IN THE GALAXY}

\begin{document}

\title{ Propagation of cosmic-ray nucleons in the Galaxy}

\author{Andrew W.~Strong\altaffilmark{1} and 
   Igor V.~Moskalenko\altaffilmark{1,2}}

\affil{\altaffilmark{1}Max-Planck-Institut f\"ur extraterrestrische Physik,
   Postfach 1603, D-85740 Garching, Germany}
\affil{\altaffilmark{2}Institute for Nuclear Physics, 
   M.V.Lomonosov Moscow State University, 119 899 Moscow, Russia}

\authoremail{aws@mpe.mpg.de; imos@mpe.mpg.de}

\begin{abstract}
We describe a method for the numerical computation of the propagation
of primary and secondary nucleons, primary electrons, and secondary
positrons and electrons.  Fragmentation and energy losses are computed
using realistic distributions for the interstellar gas and radiation
fields, and diffusive reacceleration is also incorporated.  The models
are adjusted to agree with the observed cosmic-ray \BC\ and \Berat\
ratios. Models with diffusion and convection do not account well for
the observed energy dependence of \BC, while models with reacceleration
reproduce this easily. The height of the halo propagation region is
determined, using recent \Berat\ measurements, as $>4$ kpc for
diffusion/convection models and 4 -- 12 kpc for reacceleration
models.  For convection models we set an upper limit on the velocity
gradient of $dV/dz < 7$ km s$^{-1}$ kpc$^{-1}$.  The radial
distribution of cosmic-ray sources required is broader than current
estimates of the SNR distribution for all halo sizes.  Full details of
the numerical method used to solve the cosmic-ray propagation equation
are given.

\end{abstract}

\keywords{cosmic rays --- diffusion --- elementary particles --- 
Galaxy: general --- ISM: abundances --- ISM: general}

\section{Introduction \label{introduction}}

A numerical method and corresponding computer code for the calculation
of Galactic cosmic-ray propagation has been developed, which is a
further development of the approach described by Strong \& Youssefi
(1995) and Strong (1996).  Primary and secondary nucleons, primary and
secondary electrons, and secondary positrons are included. The basic
spatial propagation mechanisms are (momentum-dependent) diffusion and
convection, while in momentum space energy loss and diffusive
reacceleration are treated.  Fragmentation and energy losses are
computed using realistic distributions for the interstellar gas and
radiation fields.  Preliminary results were presented in Strong \&
Moskalenko (1997) (hereafter \cite{StrongMoskalenko97}) and full
results for protons, Helium, positrons, and electrons in Moskalenko \&
Strong (1998a) (hereafter \cite{MoskalenkoStrong98a}).  In
\cite{MoskalenkoStrong98a} we referred the description of the numerical
method to the present paper (Paper III), and full details are now
given. Results for gamma-rays and synchrotron radiation will be given
in Moskalenko \& Strong (1998b) (hereafter
\cite{MoskalenkoStrong98b}).

We note that our positron predictions from \cite{MoskalenkoStrong98a}
have been compared with more recent absolute measurements in Barwick et
al. (1998) and the agreement is good; for the positrons this new
comparison has the advantage of being independent of the electron
spectrum, unlike the positron/electron ratio which was the main focus
of \cite{MoskalenkoStrong98a}.  The ultimate goal is to combine all
constraints including gamma-ray and synchrotron spectra; this will be
pursued in \cite{MoskalenkoStrong98b}.

The rationale for our approach was given previously
(\cite{StrongMoskalenko97,MoskalenkoStrong98a}).  Briefly, the idea is
to develop a model which simultaneously reproduces observational data
of many kinds related to cosmic-ray origin and propagation: directly
via measurements of nuclei, electrons, and positrons, indirectly via
gamma rays and synchrotron radiation.  These data provide many
independent constraints on any model and our approach is able to take
advantage of this since it must be consistent with all types of
observation.  We emphasize also the use of realistic astrophysical
input (e.g. for the gas distribution) as well as theoretical
developments (e.g. reacceleration).  The code is sufficiently general
that new physical effects can be introduced as required. We aim for a
`standard model' which can be improved with new astrophysical input and
additional observational constraints. For interested users our model is
available in the public domain on the World Wide Web.

It was pointed out many years ago (see
\cite{Ginzburg80,Berezinskii90}) that the interpretation of radioactive
cosmic-ray nuclei is model-dependent and in particular that halo models
lead to a quite different physical picture from homogeneous models.
The latter show simply a rather lower average matter density than the
local Galactic hydrogen (e.g., \cite{SimpsonGarcia88,Lukasiak94a}), but
do not lead to a meaningful estimate of the size of the confinement
region, and the correponding cosmic-ray `lifetime' is model-dependent.
In such treatments the lifetime is combined with the grammage to yield
an `average density'.  For example Lukasiak et al. (1994a) find an
`average density' of 0.28 cm$^{-3}$ compared to the local interstellar
value of about 1 cm$^{-3}$, indicating a $z$-extent of less than 1 kpc
compared to the several kpc found in diffusive halo models.
In the present work we use a model which includes spatial dimensions
as a basic element, and so these issues are automatically addressed.

The possible r\^ole of convection was shown by Jokipii (1976), and
Jones (1979) pointed out its effect on the energy-dependence of the
secondary/primary ratio.  Recent papers give estimates for the halo
size and limits on convection based on existing calculations (e.g.,
\cite{Webber92}), and in the present work we attempt to improve on
these models with a more detailed treatment.

Previous approaches to the spatial nucleon propagation problem have
been mainly analytical: Jones (1979), Freedman et al. (1980),
Berezinskii et al. (1990), Webber, Lee \& Gupta (1992) and Bloemen et
al. (1993) treated diffusion/convection models in this way.  A problem
here is that energy losses are difficult to treat and in fact were
apparently not included except by Webber, Lee \& Gupta (1992),
however even there not {\it explicitly}.  Bloemen et al. (1993) used
the `grammage' formulation rather than the explicit isotope ratios, and
their propagation equation implicitly assumes identical distributions
of primary and secondary source functions.  These papers did not
attempt to fit the low-energy ($<1$ GeV/nucleon) \BC\ data (which we
will show leads to problems) and also did not consider reacceleration.
It is clear than an analytical treatment quickly becomes limited as
soon as more realistic models are desired, and this is the main
justification for the numerical approach presented in this paper. The
case of electrons and positrons is even more intractable analytically,
although fairly general cases have been treated (\cite{Lerche82}).
Owens \& Jokipii (1977a,b) adopted an alternative approach with
Monte-Carlo simulations, for both nucleons and electrons. Recently
Porter \& Protheroe (1997) made use of this method for electrons. Both
these applications are for 1-D propagation, in the $z$-direction only.
This method allows realistic models to be computed, but would be very
time-consuming for 2- or 3-D cases. Our method, using numerical
solution of the propagation equation, is a practical alternative.
Since most of these studies were done, the data on both stable and
radioactive nuclei has improved considerably and thus a re-evaluation
is warranted.

Reacceleration has previously been handled using leaky-box calculations
(\cite{Letaw93,SeoPtuskin94,HeinbachSimon95}); this has the advantage
of allowing a full reaction network to be used (far beyond what is
possible in the present approach), but suffers from the usual
limitations of leaky-box models, especially concerning radioactive
nuclei, which were not included in these treatments.  Our simplified
reaction network is necessary because of the added spatial dimensions,
but we believe it is fully sufficient for our purpose, since we are not
attempting to derive a comprehensive isotopic composition.  A similar
approach was followed by Webber, Lee \& Gupta (1992). A more
complex reaction scheme would not in any way change our conclusions.

We model convection in a simple way, taking a linear increase of
velocity with $z$.  Detailed self-consistent models of cosmic-ray
driven MHD winds (\cite{Zirakashvili96,Ptuskin97}) provide explicit
predictions for the convective transport of cosmic-rays, and our
approach could be used in future to evaluate the observational
consequences of such models.

In this paper we concentrate on the evaluation of the \BC\ and \Berat\
ratios, evaluation of diffusion/convection and reacceleration models,
and on setting limits on the halo size.  The \BC\ data is used since
it is the most accurately measured ratio covering a wide energy range
and having well established cross sections.  The \Berat\ ratio is used
rather than $^{10}Be/(^7Be +\, ^9Be)$ since it is less sensitive to
solar modulation and to rigidity effects in the propagation.  A
re-evaluation of the halo size is desirable since new \Berat\ data are
now available from Ulysses (\cite{Connell98}) with better statistics
than previously.
It is not the purpose of this approach to perform detailed source
abundance calculations with a large network of reactions, which is
still best done with the path-length distribution approach
(\cite{DuVernois96} and references therein). Instead we use just the
principal progenitors and weighted cross sections based on the observed
cosmic-ray abundances (see \cite{Webber92}).  Other key cosmic-ray
ratios such as $^{26}Al/\,^{27}Al$ and sub-$Fe/Fe$ are beyond the scope
of this paper but will be addressed in future work.

Also important are cosmic-ray gradients as derived from gamma rays;
this provides a consistency check on the distribution of cosmic-ray
sources, and we address this here.

\section{Description of the models \label{Description}}

The models are three dimensional with cylindrical symmetry in the
Galaxy, and the basic coordinates are $(R,z,p)$, where $R$ is
Galactocentric radius, $z$ is the distance from the Galactic plane, and
$p$ is the total particle momentum.  The distance from the Sun to the
Galactic centre is taken as 8.5 kpc.  In the models the propagation
region is bounded by $R=R_h$, $z=z_h$ beyond which free escape is
assumed. We take $R_h=30$ kpc. The range $z_h=1-20$ kpc is considered
since this is suggested by previous studies of radioactive nuclei
(e.g., \cite{Lukasiak94a}) and the distribution of synchrotron
radiation (\cite{Phillipps81}).  For a given $z_h$ the diffusion
coefficient as a function of momentum is determined by \BC\ for the
case of no reacceleration; if reacceleration is assumed then the
reacceleration strength (related to the Alfv\'en speed) is constrained
by the energy-dependence of \BC.  The spatial diffusion coefficient for
the case of no reacceleration is taken as $\Dxx = \beta
D_0(\rho/\rho_0)^{\delta_1}$ below rigidity $\rho_0$, $\beta
D_0(\rho/\rho_0)^{\delta_2}$ above rigidity $\rho_0$, where the factor
$\beta$ ($= v/c$) is a natural consequence of a random-walk process.
Since the introduction of a sharp break in $\Dxx$ is an extremely
contrived procedure which is adopted just to fit \BC\ at all energies,
we also consider the case $\delta_1=\delta_2$, i.e. no break, in order
to investigate the possibility of reproducing the data in a physically
simpler way\footnote{ In \cite{MoskalenkoStrong98a} we considered only
$\delta_1 = 0$ and did not consider convection}.  The convection
velocity (in $z$-direction only) $V(z)$ is assumed to increase linearly
with distance from the plane ($V>0$ for $z>0$, $V<0$ for $z<0$, and
$dV/dz>0$ for all $z$).  This implies a constant adiabatic energy loss;
the possibility of adiabatic energy gain ($dV/dz < 0$) is not
considered. The linear form for $V(z)$ is consistent with cosmic-ray
driven MHD wind models (e.g., \cite{Zirakashvili96}).  The velocity at
$z = 0$ is a model parameter, but we consider here only $V(0) = 0$.

Some stochastic reacceleration is inevitable, and it provides a natural
mechanism to reproduce the energy dependence of the \BC\ ratio without
an {\it ad hoc} form for the diffusion coefficient
(\cite{Letaw93,SeoPtuskin94,HeinbachSimon95,SimonHeinbach96}). 
The spatial diffusion coefficient for the case of reacceleration
assumes a Kolmogorov spectrum of weak MHD turbulence so $\Dxx=\beta
D_0(\rho/\rho_0)^\delta$ with $\delta=1/3$ for all rigidities. Simon
and Heinbach (1995) showed that the Kolmogorov form best reproduces the
observed \BC\ variation with energy.  For the case of reacceleration
the momentum-space diffusion coefficient $D_{pp}$ is related to the
spatial coefficient using the formula given by Seo \& Ptuskin (1994)
(their equation [9]), and Berezinskii et al.\ (1990)
\begin{equation}
\label{2.1}
\Dpp\Dxx = {4 p^2 {v_A}^2\over 3\delta(4-\delta^2)(4-\delta) w}\ ,
\end{equation}
where $w$ characterises the level of turbulence, and is equal to the ratio
of MHD wave energy density to magnetic field energy density. 
The main free parameter in this relation is the Alfv\'en speed $v_A$;
we take $w = 1$ (\cite{SeoPtuskin94}) but clearly only the quantity
$v_A^2 /w$ is relevant. 

The atomic hydrogen distribution is represented by the formula
\begin{equation}
\label{2.3}
n_{HI}(R,z)=n_{HI}(R) e^{ -\ln2 \cdot (z/z_0)^2} ,
\end{equation}
where $n_{HI}(R)$ is taken from Gordon \& Burton (1976) 
and $z_0$ from Cox, Kr\"ugel \& Mezger (1986) giving an exponential
increase in the width of the $HI$ layer outside the solar circle:

\begin{equation}
\label{2.4}
z_0(R) = \left\{
\begin{array}{l l}
0.25{\rm\ kpc}, &\ R \le 10{\rm\ kpc}; \\
0.083\, e^{0.11R}{\rm\ kpc}, &\ R > 10{\rm\ kpc}.
\end{array} \right.
\end{equation}
The distribution of molecular hydrogen is taken from Bronfman et al.
(1988) using $CO$ surveys:
\begin{equation}
\label{2.5}
n_{H_2}(R,z)=n_{H_2}(R)\, e^{ -\ln2 \cdot (z/70 {\rm\ pc})^2}\ . 
\end{equation}
The adopted radial distribution of $HI$ and $H_2$ is shown in
Figure~\ref{fig1}.

For the ionized gas we use the two-component model of Cordes et al.
(1991):
\begin{equation}
\label{2.6}
n_{HII}=0.025\, e^{ -{|z|\over 1{\rm\ kpc}}
-\left({R\over 20{\rm\ kpc}}\right)^2}
+ 0.2\, e^{ -{|z|\over 0.15{\rm\ kpc}}
-\left( {R\over 2{\rm\ kpc}} -2 \right)^2} 
{\rm\ cm}^{-3}\ .
\end{equation}
The first term represents the extensive warm ionized gas and is similar
to the distribution given by Reynolds (1989); the second term
represents $HII$ regions and is concentrated around $R = 4$ kpc.  A
temperature of $10^{4}$ K is assumed to compute Coulomb energy losses
in ionized gas.

\medskip
\centerline{\includegraphics[width=\fwc]{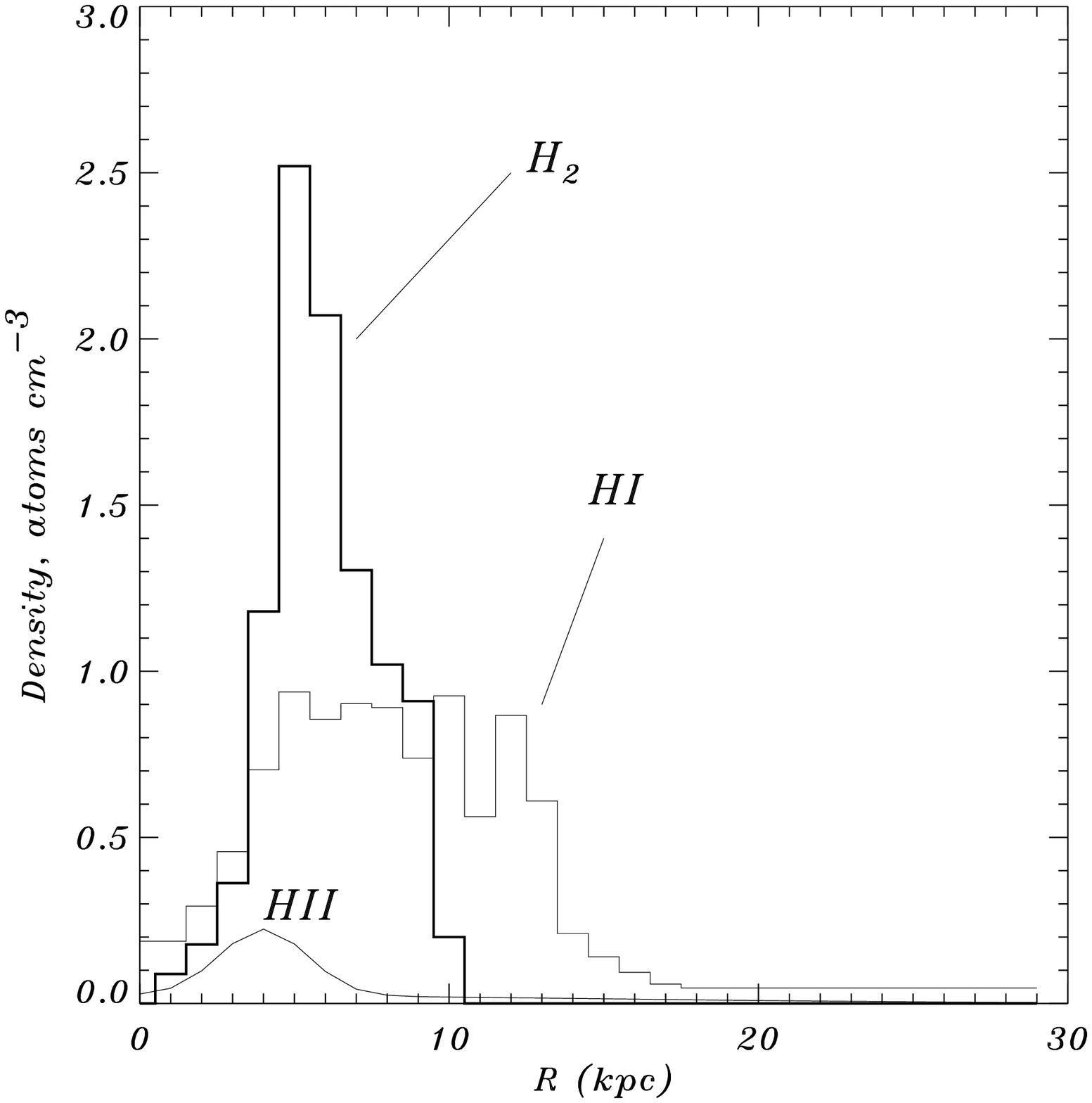}}
\vskip -5pt
\figcaption[fig1.ps]{ The adopted radial distribution of atomic,
molecular and ionized hydrogen at $z = 0$.  \label{fig1} }
\medskip

\begin{figure*}[hbt]
\hfill\includegraphics[width=\fwb]{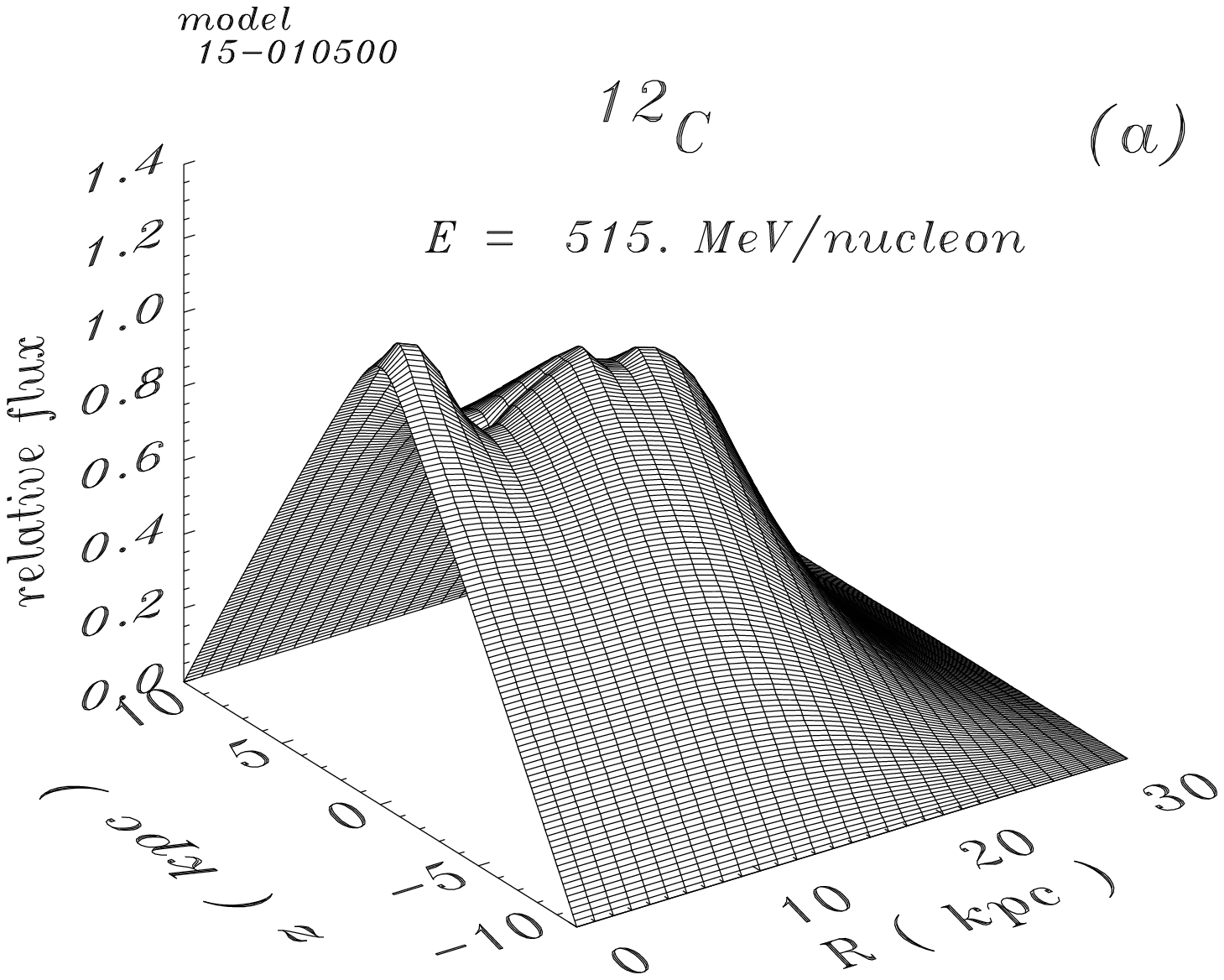}\hfill
      \includegraphics[width=\fwb]{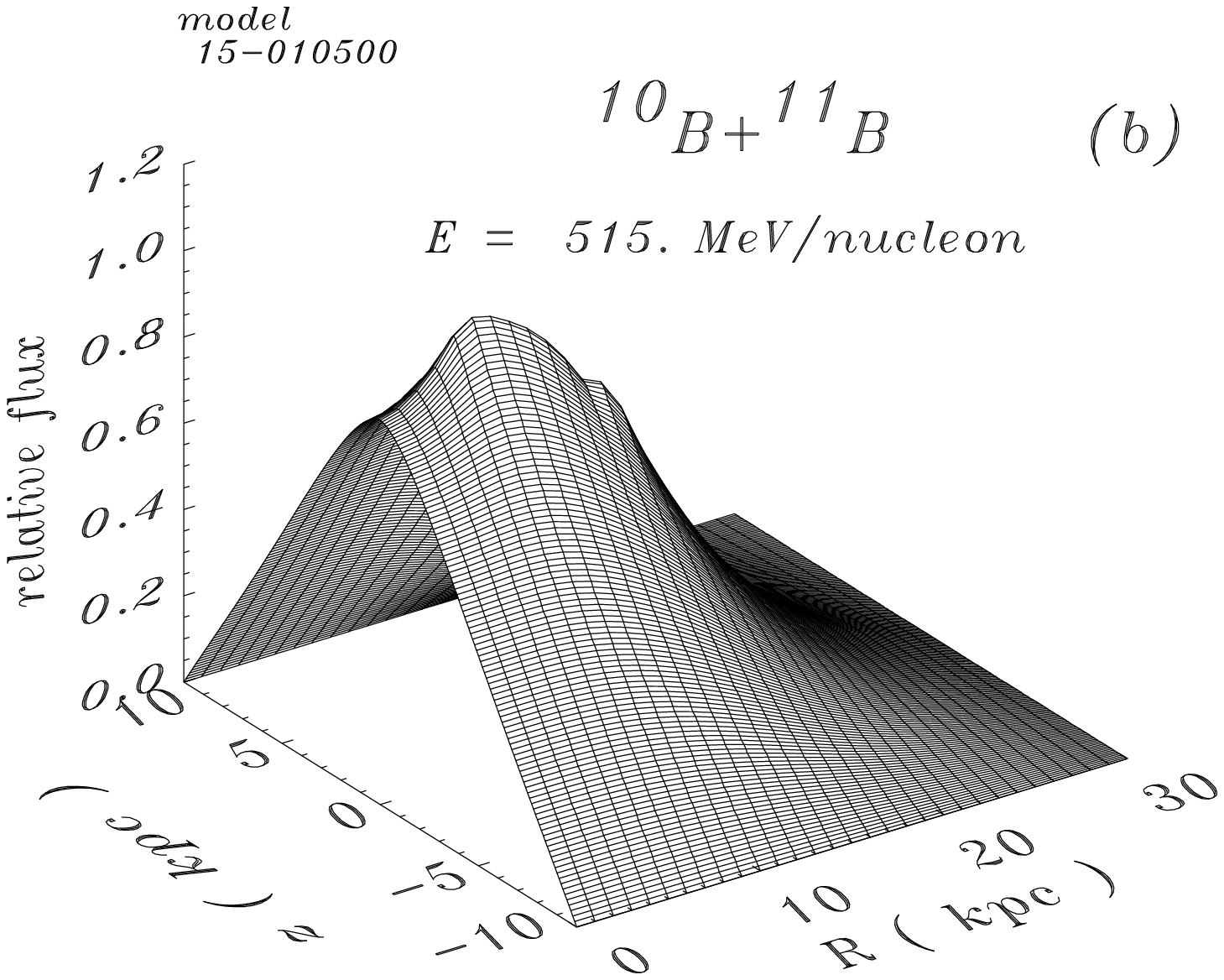}\hfill
\vskip -5pt
\caption[fig2a.ps,fig2b.ps]{ \footnotesize
The 3-D distribution of $^{12}C$ and
$^{10,11}B$ at 515 MeV/nucleon for reacceleration model with $z_h$ = 10
kpc, for $v_A$ = 20 km s$^{-1}$.  Parameters: see model 10500 in
Table~\ref{table2}.  \label{fig2} }
\end{figure*}

The $He/H$ ratio of the interstellar gas is taken as 0.11 by number;
there is some uncertainty in this quantity, but our value is consistent
with recent photospheric determinations ($0.10\pm0.008$:
\cite{Grevesse}).  Helioseismological methods (\cite{Hernandez94}) give
a Helium abundance by mass of $0.242$ corresponding to $He/H = 0.08$,
but still with possible uncertainties due to the details of the
models.  Although the latter is perhaps the most accurate local
determination, the uncertainty in extending the photospheric value to
the interstellar medium over the whole Galaxy is large.  Other
uncertainties dominate the secondary production, for example the
density of neutral and molecular hydrogen.  In any case, even if $He/H
= 0.08$ the influence of the uncertainty of $He/H$ on the secondary
production does not exceed 10\%.

The distribution of cosmic-ray sources is chosen to reproduce (after
propagation) the cosmic-ray distribution determined by analysis of
EGRET gamma-ray data (\cite{StrongMattox96}).  The form used is
\begin{equation}
\label{2.2}
q(R,z) = q_0 \left({R\over R_\odot}\right)^\eta 
e^{-\xi{R-R_\odot\over R_\odot} -{|z|\over 0.2{\rm\ kpc}}}\ ,
\end{equation}
where $q_0$ is a normalization constant, $\eta$ and $\xi$ are 
parameters; the $R$-dependence has the same parameterization as
that used for SNR by Case \& Bhattacharya (1996), but we adopt
different parameters in order to fit the gamma-ray gradient.  We also
compute models with the SNR distribution, to investigate the
possibility of fitting the gradient in this case.  We apply a cutoff in
the source distribution at $R = 20$ kpc since it is unlikely that
significant sources are present at such large radii.  The
$z$-dependence of $q$ is nominal and reflects simply the assumed
confinement of sources to the disk.

We assume that the source distribution of all cosmic-ray primaries is
the same.  Meyer, Drury \& Ellison (1997) suggest that part of the C
and O originates in acceleration of C and O enriched pre-SN Wolf-Rayet
wind material by supernovae, but the source distribution in this case
would still follow that of SNR.

The primary propagation is computed first giving the primary
distribution as a function of ($R, z, p$); then the secondary source
function is obtained from the gas density and cross sections, and
finally the secondary propagation is computed.  Tertiary reactions,
such as $^{11}B\to\, ^{10}B$ are treated as described in Appendix~A.
The entire calculation is performed with momentum as the kinematic
variable, since this greatly facilitates the inclusion of
reacceleration.

Full details of the propagation equation and numerical method used are
given in Appendices~A and B.  The method encompasses nucleons,
electrons and positrons.  Energy losses for nucleons by ionization and
Coulomb interactions are included following Mannheim \& Schlickeiser
(1994) (see Appendix~C.1).  Details of the positron source function,
magnetic field and interstellar radiation field models were given in
\cite{MoskalenkoStrong98a}, and the energy loss formulae for electrons
are given in the present paper in Appendix~C.2.

As an illustration of the calculations performed by the code,
Figure~\ref{fig2} shows the $(R,z)$ distribution of primary $^{12}C$
and secondary $^{10,11}B$ at 515 MeV/nucleon for a reacceleration model
with $z_h = 10$ kpc.  In practice we are only interested in the isotope
ratios at the solar position, but it is worth noting the variations
over the Galaxy, which are due to effect of the inhomogeneous
distribution of sources and gas on the secondary production,
fragmentation and energy losses. For comparison with gamma-ray data the
full 3-D distribution is of course important and will be addressed in
\cite{MoskalenkoStrong98b}, but here only the radial cosmic-ray
gradient from gamma-rays is considered.

\section{Evaluation of models}

We consider the cases of diffusion+convection and
diffusion+reacceleration, since these are the minimum combinations
which can reproduce the key observations. In principle all three
processes could be significant, and such a general model can be
considered if independent astrophysical information or models, for
example for a Galactic wind (e.g., \cite{Zirakashvili96,Ptuskin97}),
were to be used.  Anticipating the results, it can be noted at the
outset that the reacceleration models are more satisfactory in meeting
the constraints provided by the data, reproducing the \BC\ energy
dependence without {\it ad hoc} variations in the diffusion
coefficient; further it is not possible to find any {\it simple}
version of the diffusion/convection model which reproduces
\BC\ satisfactorily without additional special assumptions.

\begin{figure*}[thb]
\hfill\includegraphics[width=\fwc]{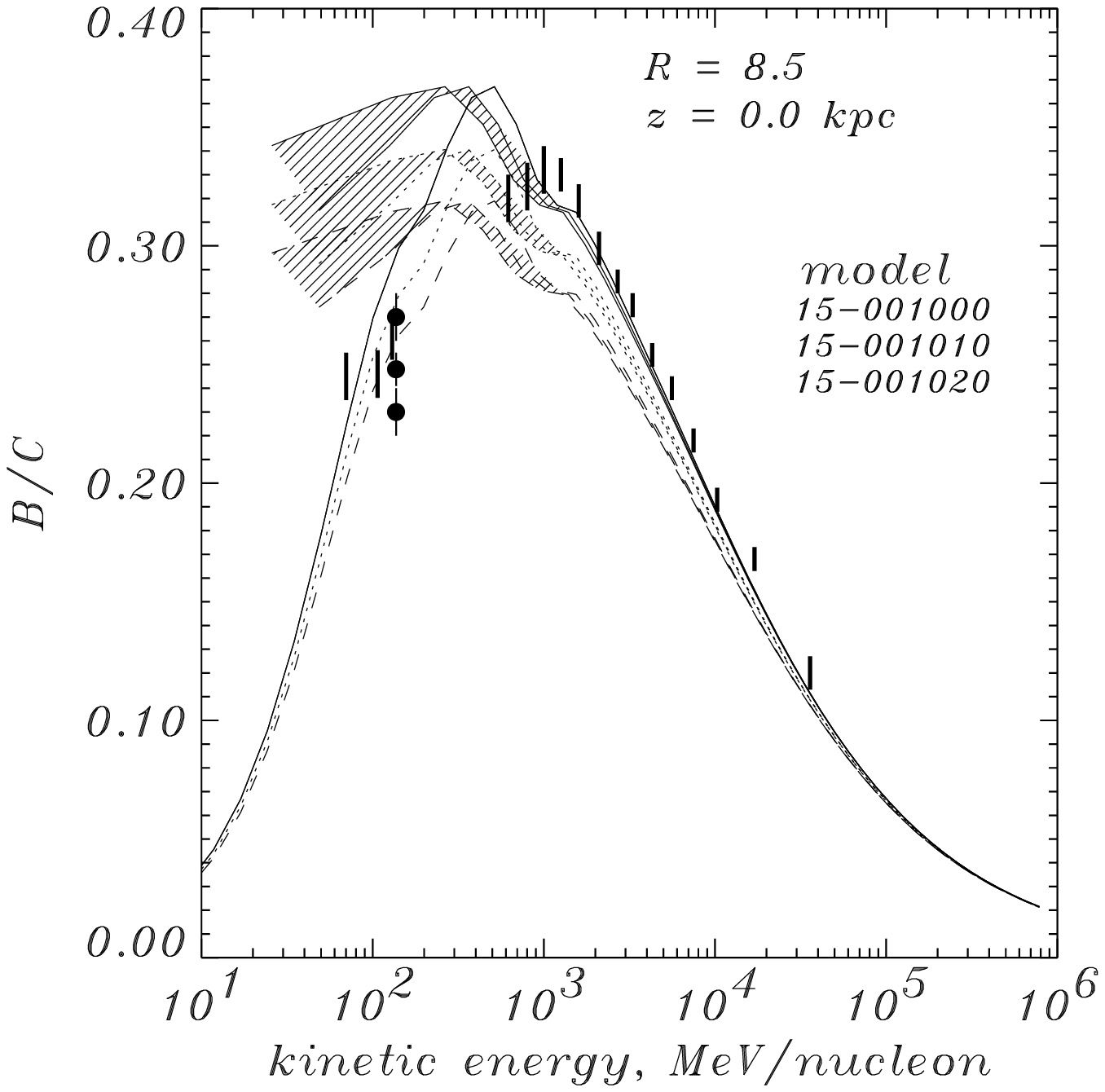}\hfill
      \includegraphics[width=\fwc]{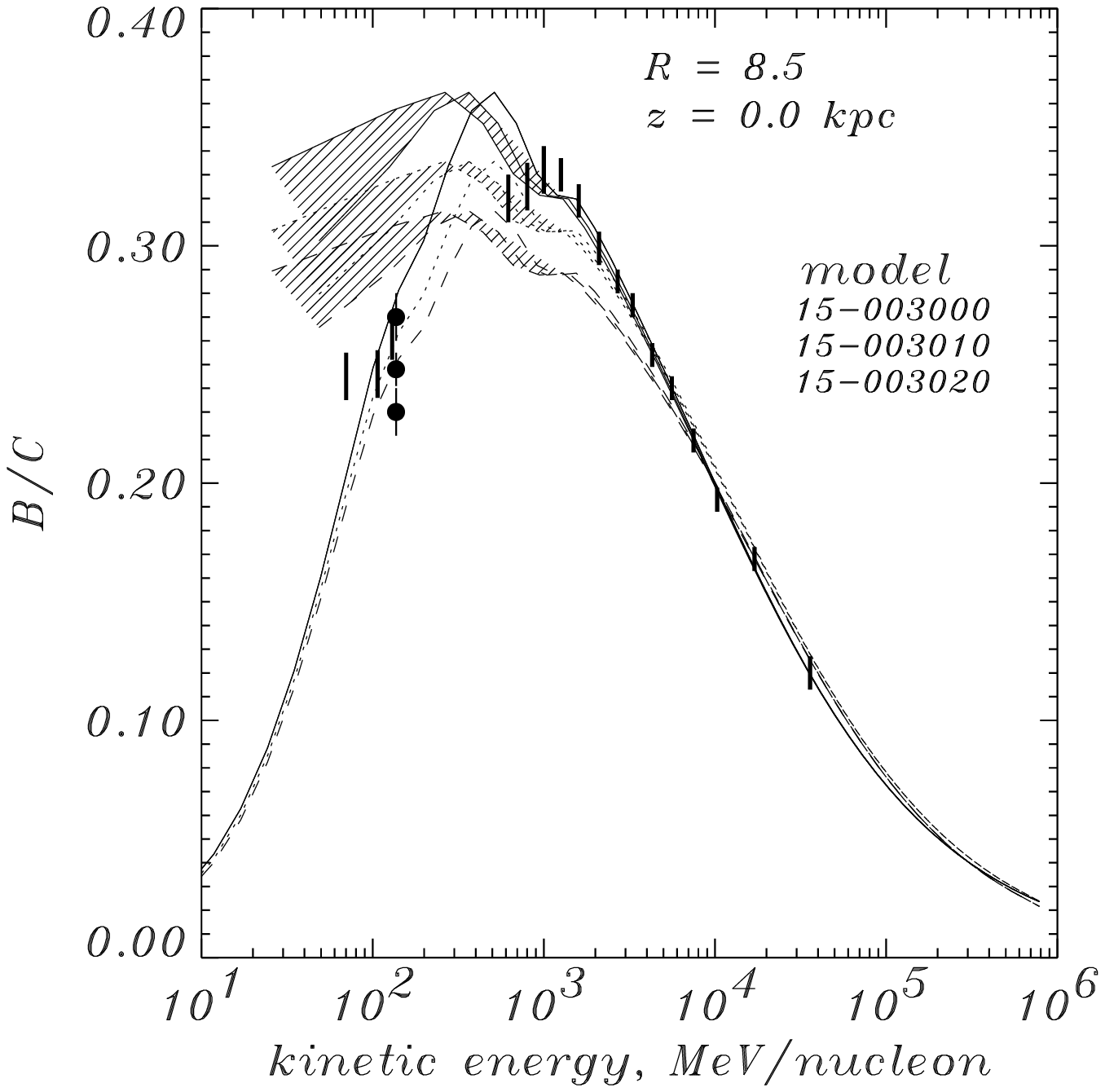}\hfill
      \includegraphics[width=\fwc]{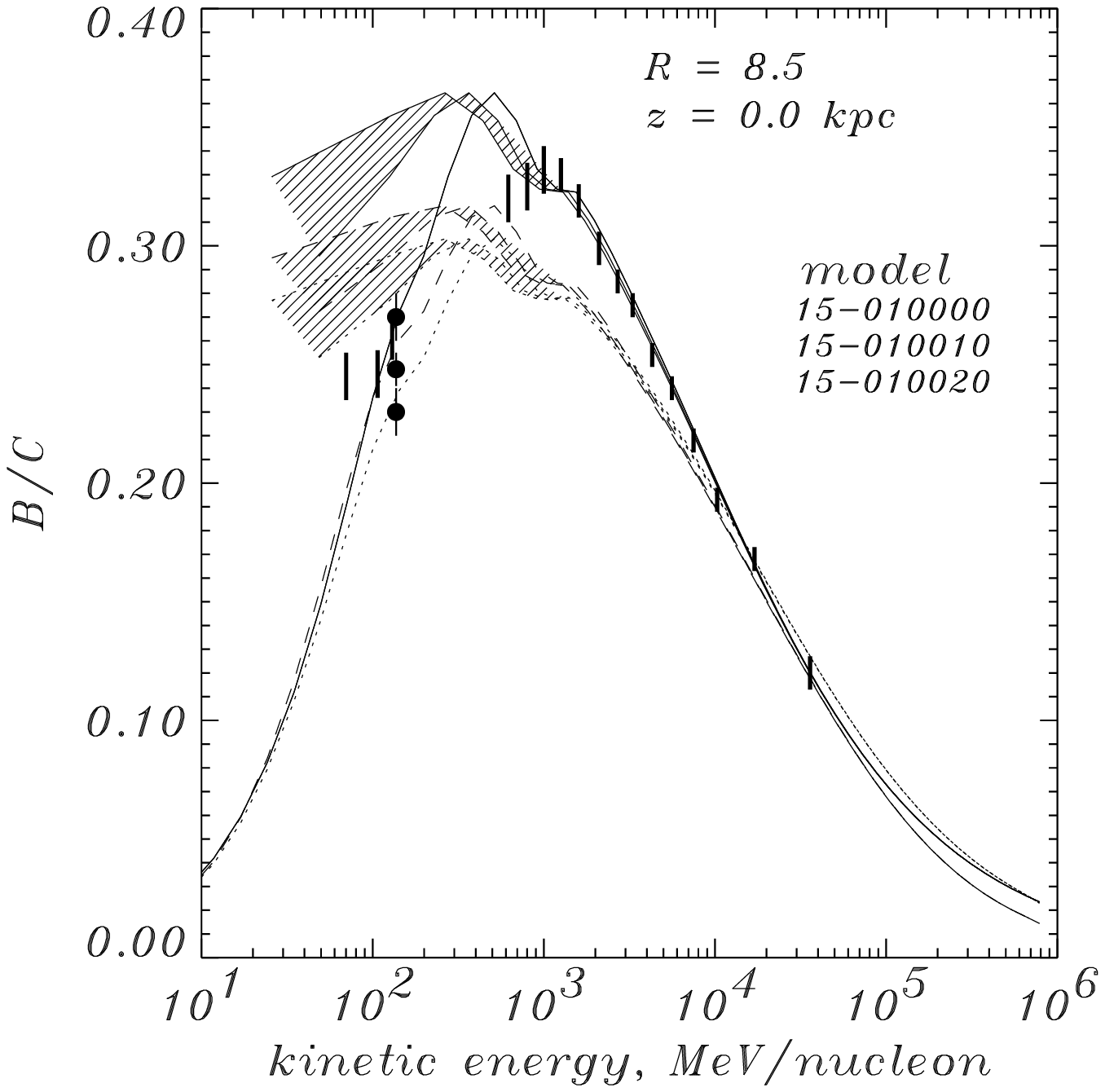}\hfill
\vskip -5pt
\caption[fig3a.ps,fig3b.ps,fig3c.ps]{ \footnotesize
\BC\ ratio for
diffusion/convection models without break in diffusion coefficient, for
$dV/dz$ = 0 (solid lines), 5 (dotted lines), and 10 km s$^{-1}$
kpc$^{-1}$ (dashed lines).  The cases shown are (a) $z_h$ = 1 kpc,
(b) $z_h$ = 3 kpc, (c) $z_h$ = 10 kpc.  Solid lines: interstellar
ratio, shaded area: modulated to 300 -- 500 MV.  Data: vertical bars:
HEAO-3, Voyager (\cite{Webber96}), filled circles: Ulysses
(\cite{DuVernois96}: $\Phi$ = 600, 840, 1080 MV).  Parameters as in
Table~\ref{table1}.  \label{fig3} }
\end{figure*} 

\begin{figure*}[thb]
\hfill\includegraphics[width=\fwc]{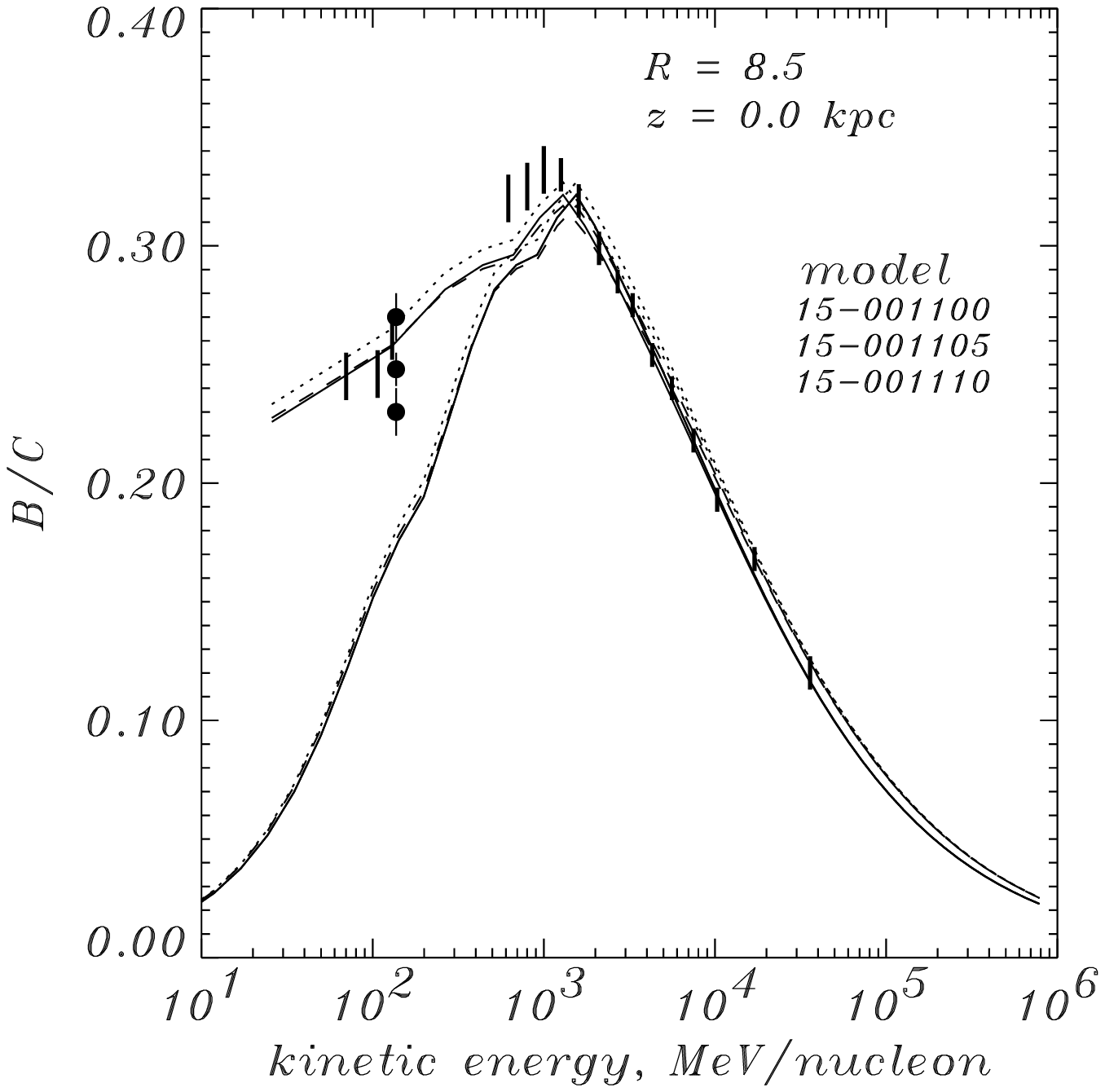}\hfill
      \includegraphics[width=\fwc]{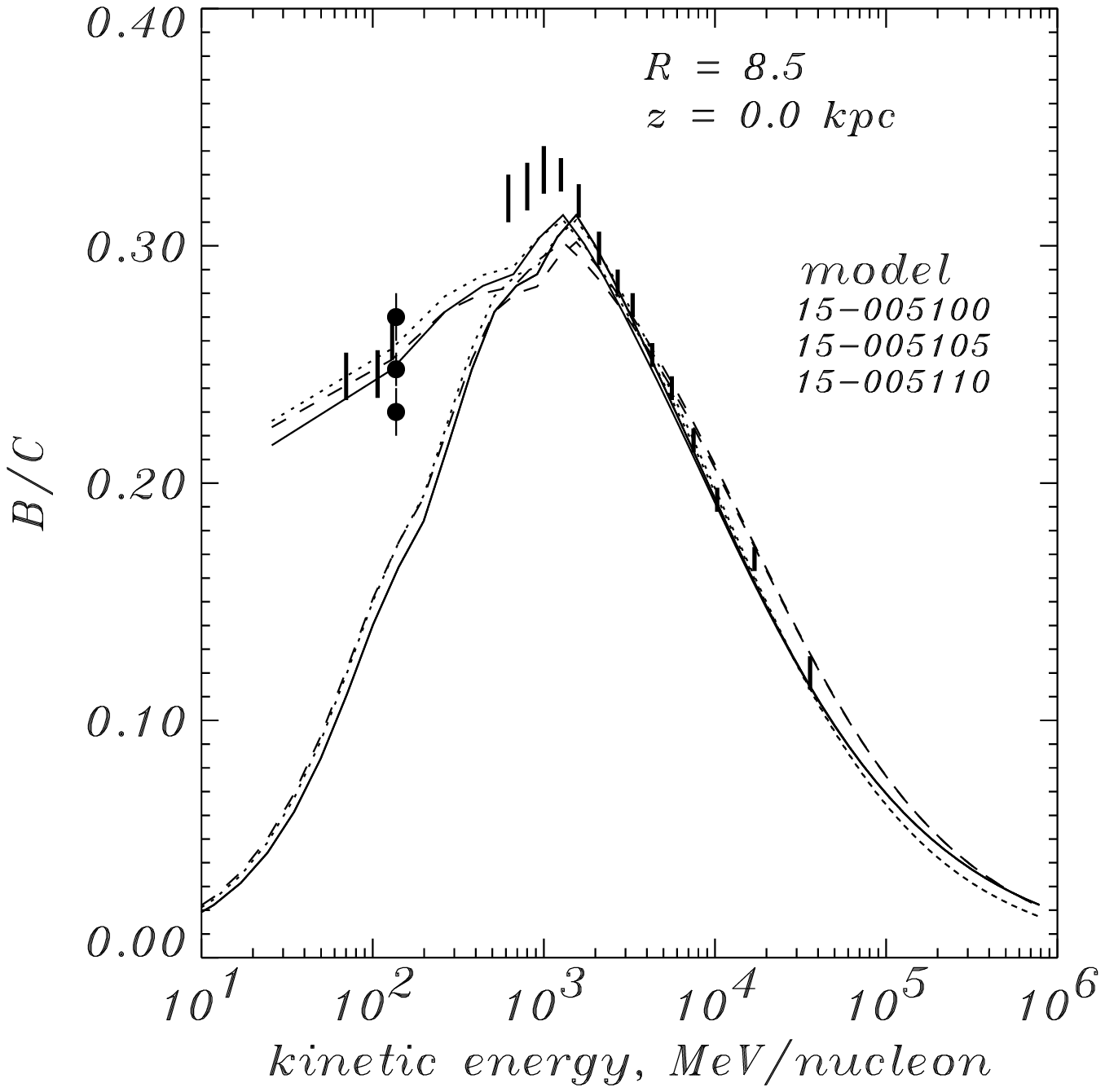}\hfill
      \includegraphics[width=\fwc]{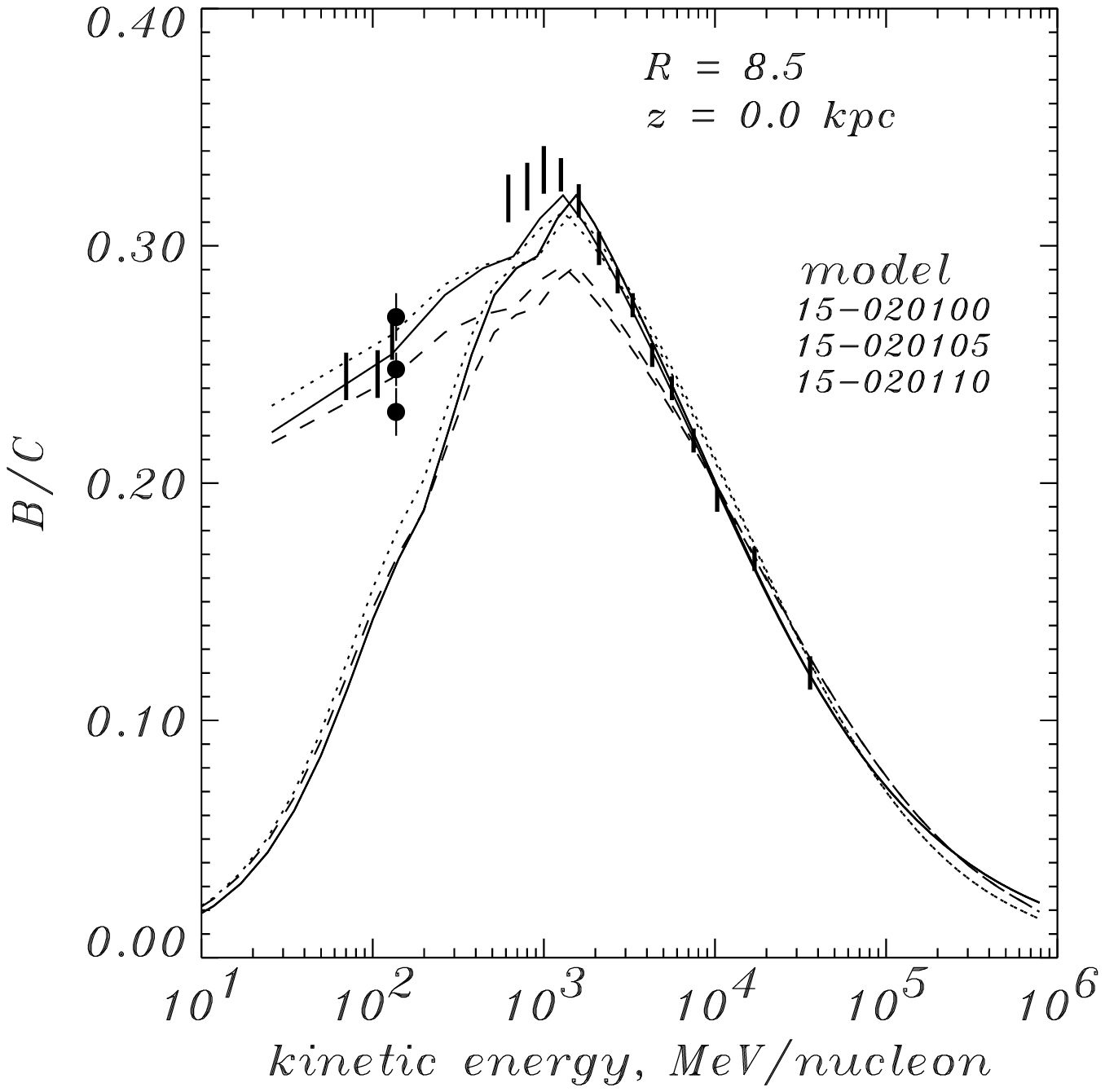}\hfill
\vskip -5pt
\caption[fig4a.ps,fig4b.ps,fig4c.ps]{  \footnotesize
\BC\ ratio for
diffusion/convection models with break in diffusion coefficient, for
$dV/dz$ = 0 (solid lines), 5 (dotted lines), and 10 km s$^{-1}$
kpc$^{-1}$ (dashed lines).  The cases shown are (a) $z_h$ = 1 kpc,
(b) $z_h$ = 5 kpc, (c) $z_h$ = 20 kpc.  Lower lines: interstellar
ratio; upper lines: modulated to 500 MV.  Parameters as in
Table~\ref{table1}.  Data: as Figure~\ref{fig3}.  \label{fig4} }
\end{figure*} 

\begin{figure*}[thb]
\hfill\includegraphics[width=\fwc]{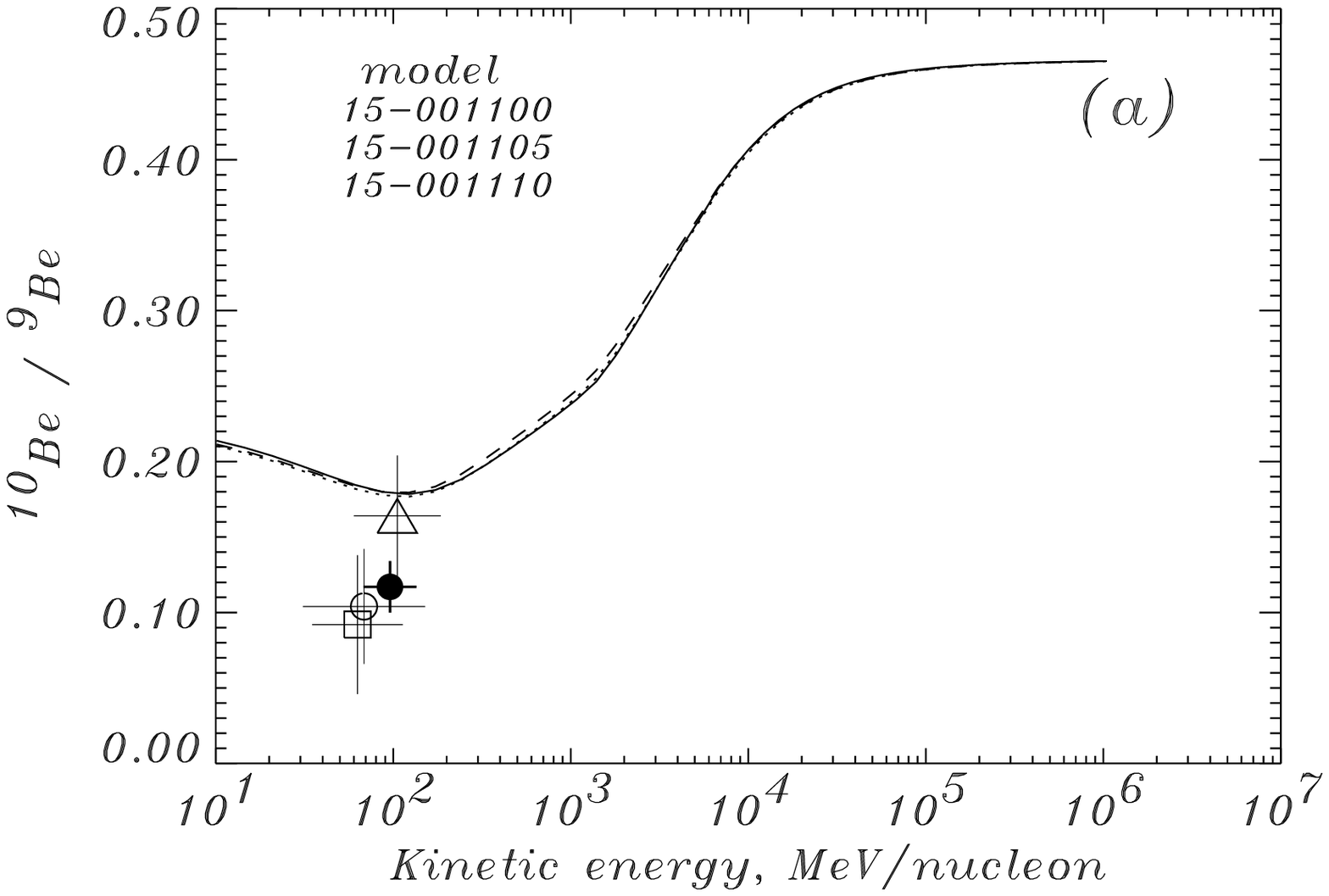}\hfill
      \includegraphics[width=\fwc]{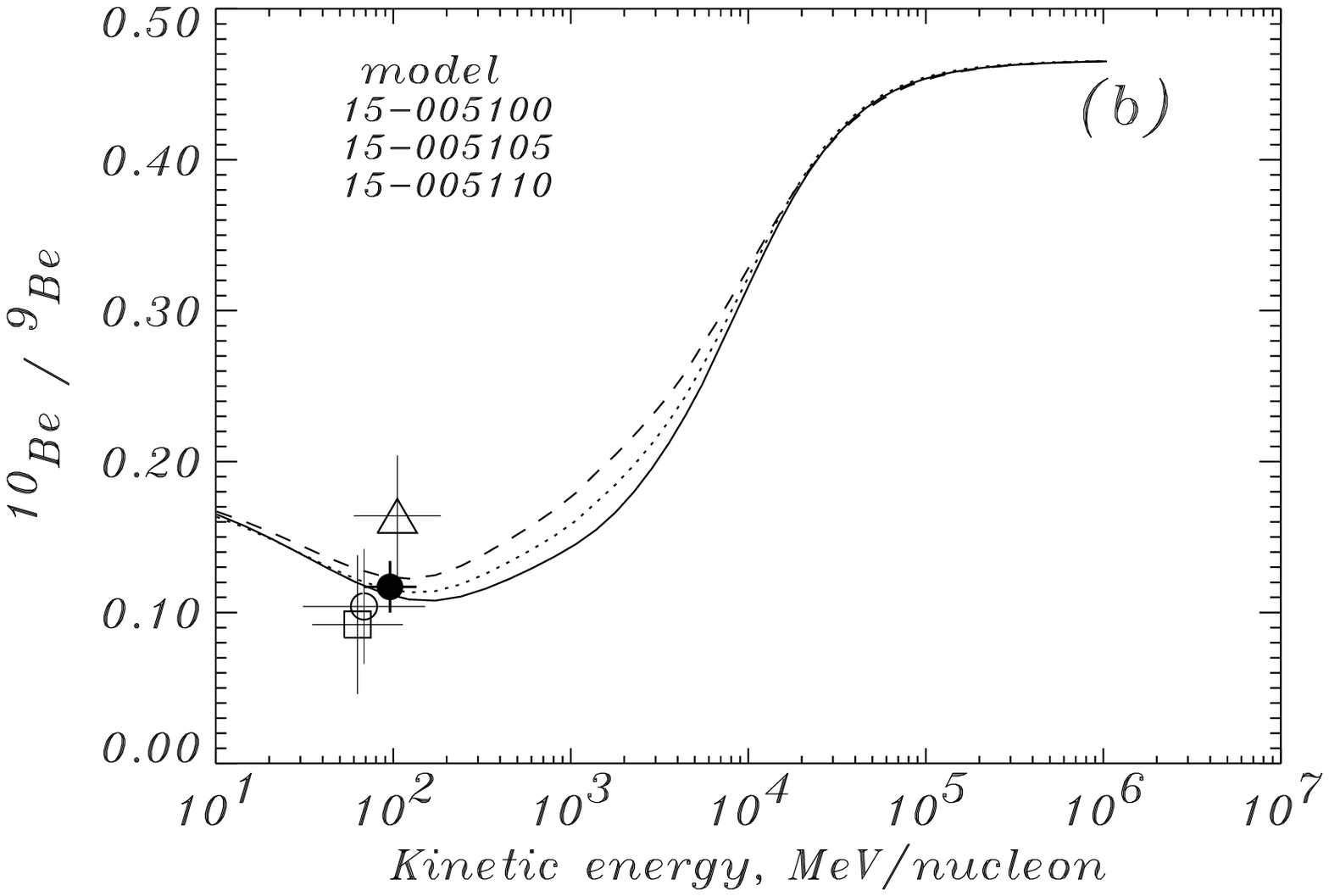}\hfill
      \includegraphics[width=\fwc]{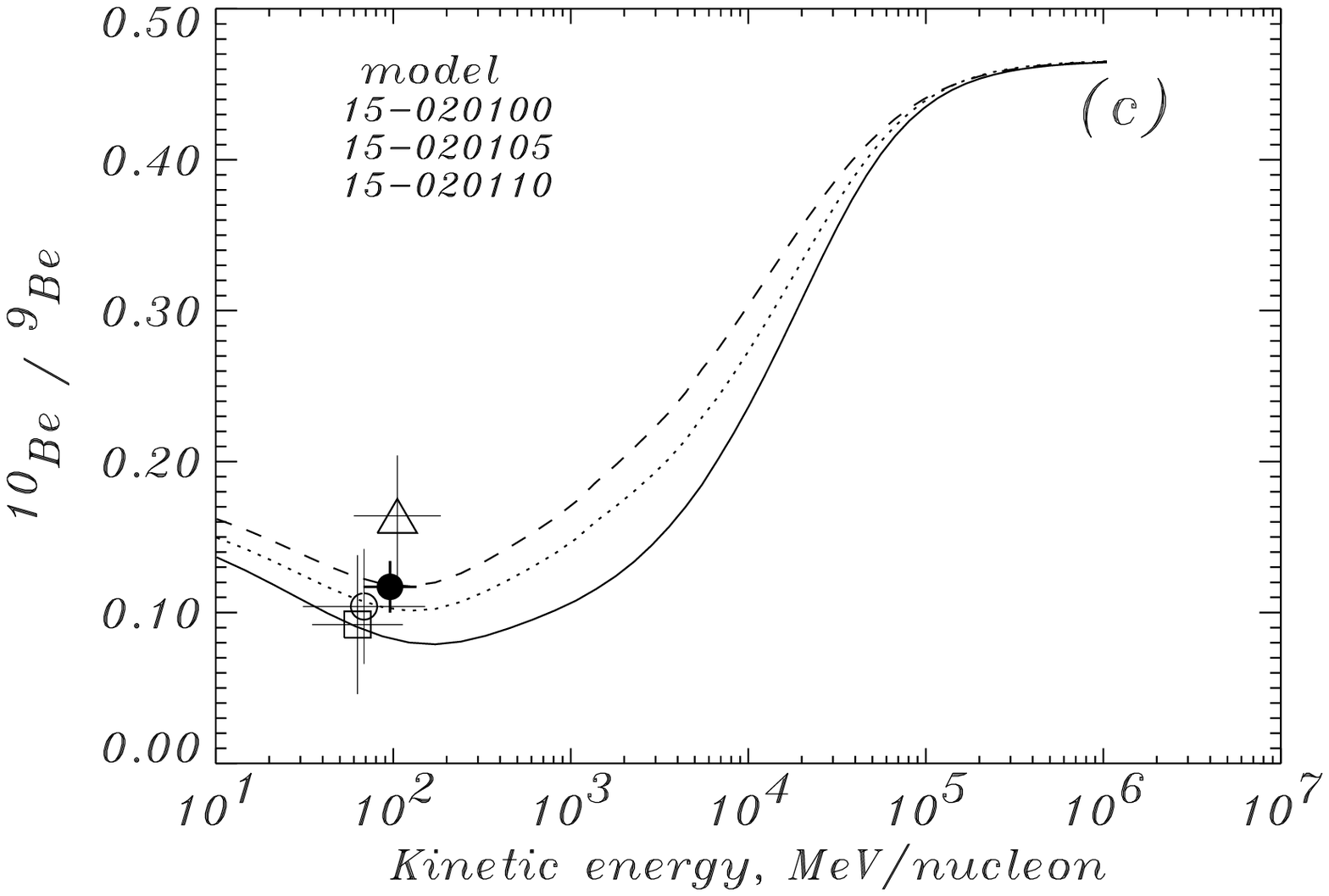}\hfill
\vskip -5pt
\caption[fig5a.ps,fig5b.ps,fig5c.ps]{ \footnotesize
\Berat\ ratio for
diffusion/convection models, for $dV/dz$ = 0 (solid lines), 5 (dotted
lines), and 10 km s$^{-1}$ kpc$^{-1}$ (dashed lines).  The cases shown
are (a) $z_h$ = 1 kpc, (b) $z_h$ = 5 kpc, (c) $z_h$ = 20 kpc.  Data
points from Lukasiak et al. (1994a) (Voyager-1,2: square, IMP-7/8: open
circle, ISEE-3: triangle) and Connell (1997) (Ulysses): filled
circle.  Parameters as in Table~\ref{table1}.  \label{fig5} }
\end{figure*} 

\begin{figure*}[thb]
\centerline{
\includegraphics[width=\fwc]{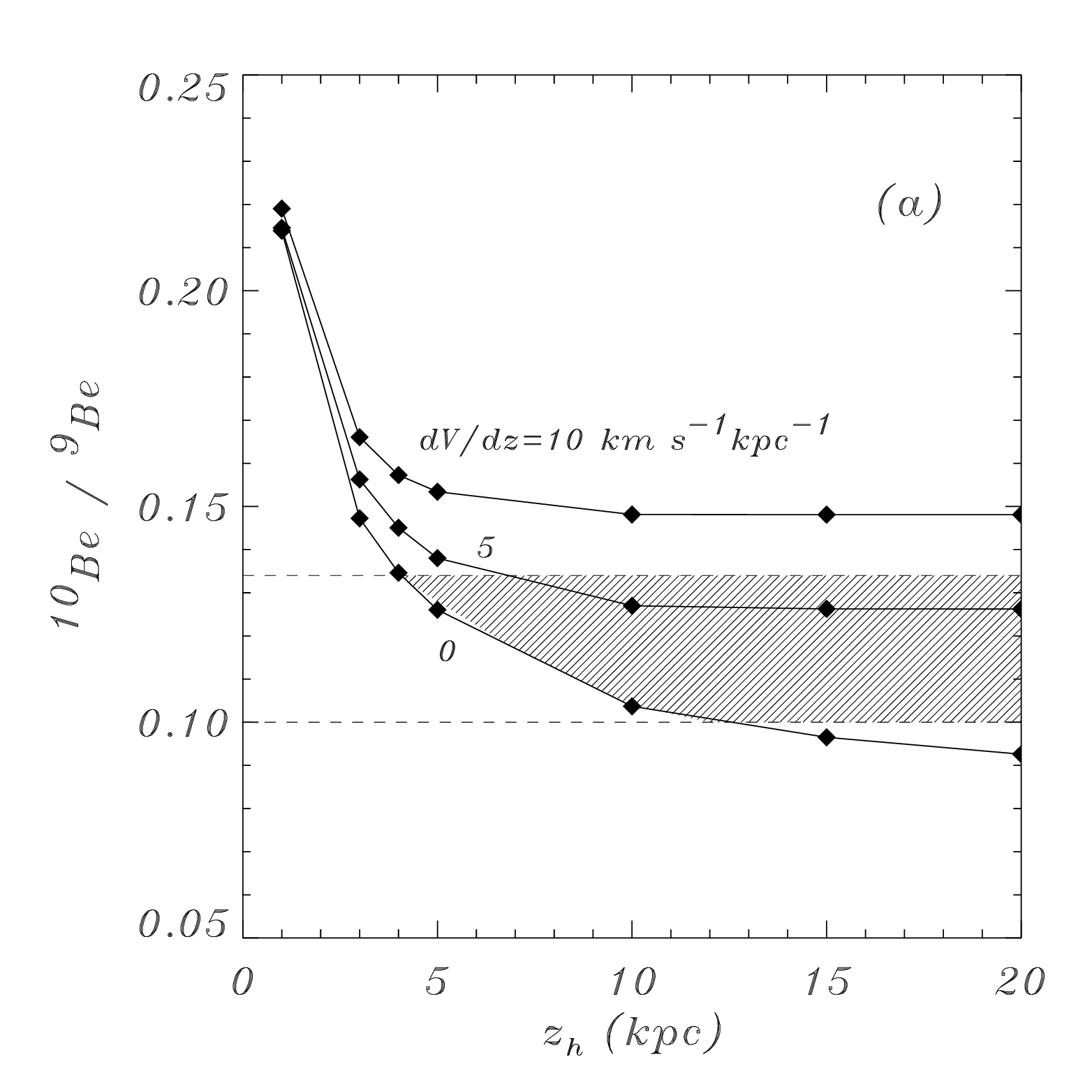}\hskip 2cm
\includegraphics[width=\fwc]{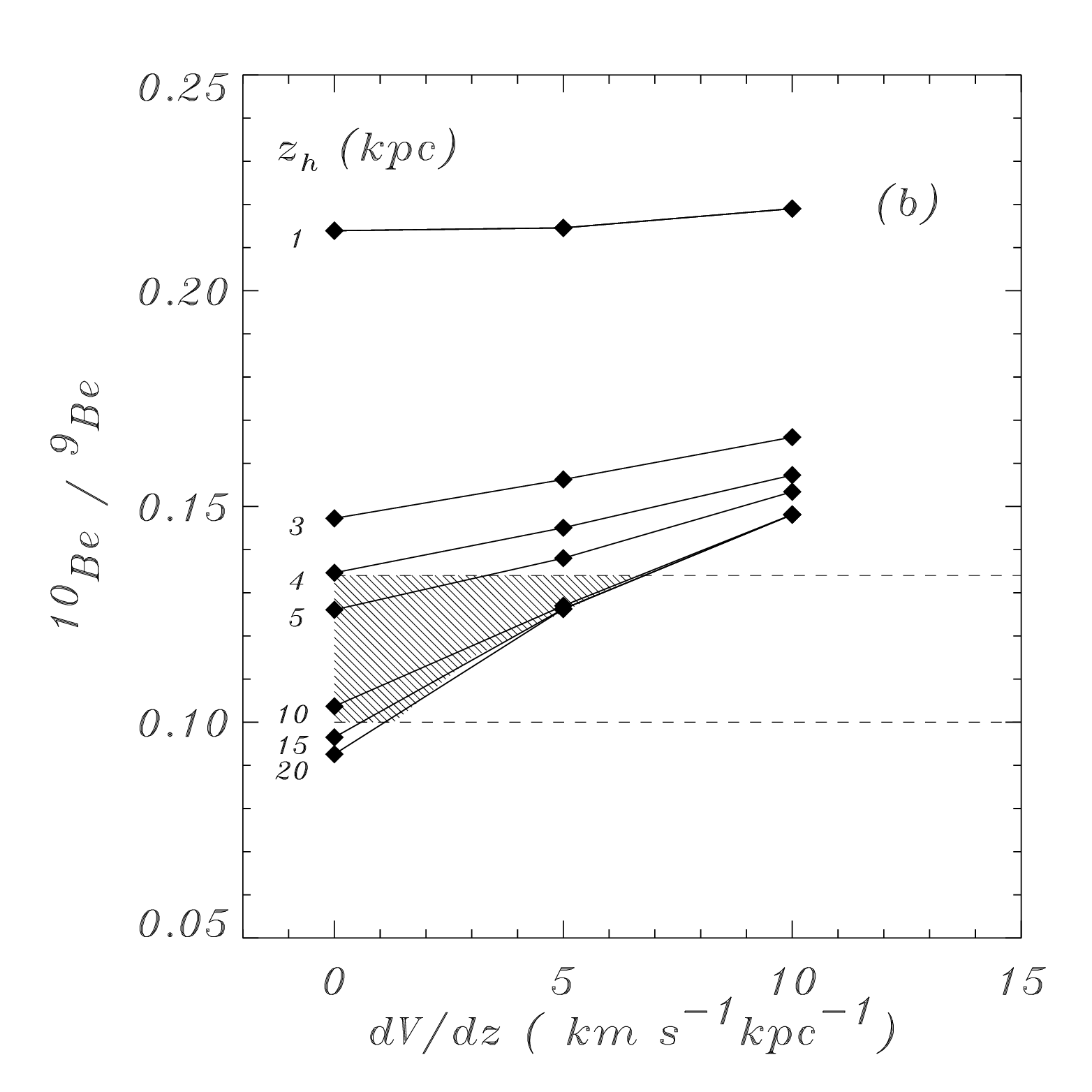}
}
\vskip -5pt
\caption[fig6a.ps,fig6b.ps]{ \footnotesize
Predicted \Berat\ ratio as function of
(a) $z_h$ for $dV/dz$ = 0, 5, 10 km s$^{-1}$ kpc$^{-1}$, (b) $dV/dz$
for $z_h = 1 - 20$ kpc at 525 MeV/nucleon corresponding to the mean
interstellar value for the Ulysses data (\cite{Connell98}); the Ulysses
experimental limits are shown as horizontal dashed lines.  The shaded
regions show the parameter ranges allowed by the data.
\label{fig6} }
\end{figure*}

In our calculations we use the \BC\ data summarized by Webber et
al.\ (1996), from HEAO--3 and Voyager 1 and 2.  The spectra were
modulated to 500 MV appropriate to this data using the force-field
approximation (\cite{GleesonAxford68}).  We also show \BC\ values from
Ulysses (\cite{DuVernois96}) for comparison, but since this has large
modulation (600 -- 1080 MV) we do not base conclusions on these
values.  We use the measured \Berat\ ratio from Ulysses
(\cite{Connell98}) and from Voyager--1,2, IMP--7/8, ISEE--3 as
summarized by Lukasiak et al. (1994a).

The source distribution adopted has $\eta=0.5$, $\xi=1.0$ in
eq.~(\ref{2.2}) (apart from the cases with SNR source distribution).
This form adequately reproduces the small observed gamma-ray based
gradient, for all $z_h$; a more detailed discussion is given in 
Section~4.

\subsection{Diffusion/convection models}
The main parameters are $z_h$, $D_0$, $\delta_1$, $\delta_2$ and
$\rho_0$ and $dV/dz$.  We treat $z_h$ as the main unknown quantity, and
consider values 1 -- 20 kpc.  The parameters of these models are
summarized in Table~\ref{table1}.  For a given $z_h$ we show \BC\ for a
series of models with different $dV/dz$.

Figure~\ref{fig3} shows the case of no break, $\delta_1 = \delta_2$;
for each $dV/dz$, the remaining parameters $D_0$, $\delta_1$ and
$\rho_0$ are adjusted to fit the data as well as possible.  It is clear
that a {\it good} fit is {\it not} possible; the basic effect of
convection is to reduce the variation of \BC\ with energy, and although
this improves the fit at low energies the characteristic peaked shape
of the measured \BC\ cannot be reproduced.  Although modulation makes
the comparison with the low energy Voyager data somewhat uncertain,
Figure~\ref{fig3} shows that the fit is unsatisfactory; the same is
true even if we use a very low modulation parameter of 300 MV in an
attempt to improve the fit.  This modulation is near the minimum value
for the entire Voyager 17 year period (cf. the average value of 500 MV;
\cite{Webber96}).  The failure to obtain a good fit is an important
conclusion since it shows that the simple inclusion of convection
cannot solve the problem of the low-energy falloff in \BC.

Since the inclusion of a convective term is nevertheless of interest
for independent astrophysical reasons (Galactic wind) we can force a
fit to the data by allowing a break in $\Dxx(p)$, with $\delta_1 \ne
\delta_2$.  Figure~\ref{fig4} shows cases with a break: here the
parameters $D_0$, $\delta_1$, $\delta_2$ and $\rho_0$ are adjusted.  In
the absence of convection, the falloff in \BC\ at low energies requires
that the diffusion coefficient increases rapidly below $\rho_0 = 3$ GV
($\delta_1\sim -0.6$) reversing the trend from higher energies
($\delta_2 \sim +0.6$).  Inclusion of the convective term does not
reduce the size of the {\it ad hoc} break in the diffusion coefficient,
in fact it rather exacerbates the problem by requiring a larger
break\footnote{Note that the dependence of interaction rate on particle
velocity itself is not sufficient to cause the full observed low-energy
falloff.  In leaky-box treatments the low-energy behaviour is modelled
by adopting a constant path-length below a few GeV/nucleon, without
attempting to justify this physically.  A convective term is often
invoked, but our treatment shows that this alone is not sufficient.}.

\begin{figure*}[t]
\centerline{
\includegraphics[width=\fwc]{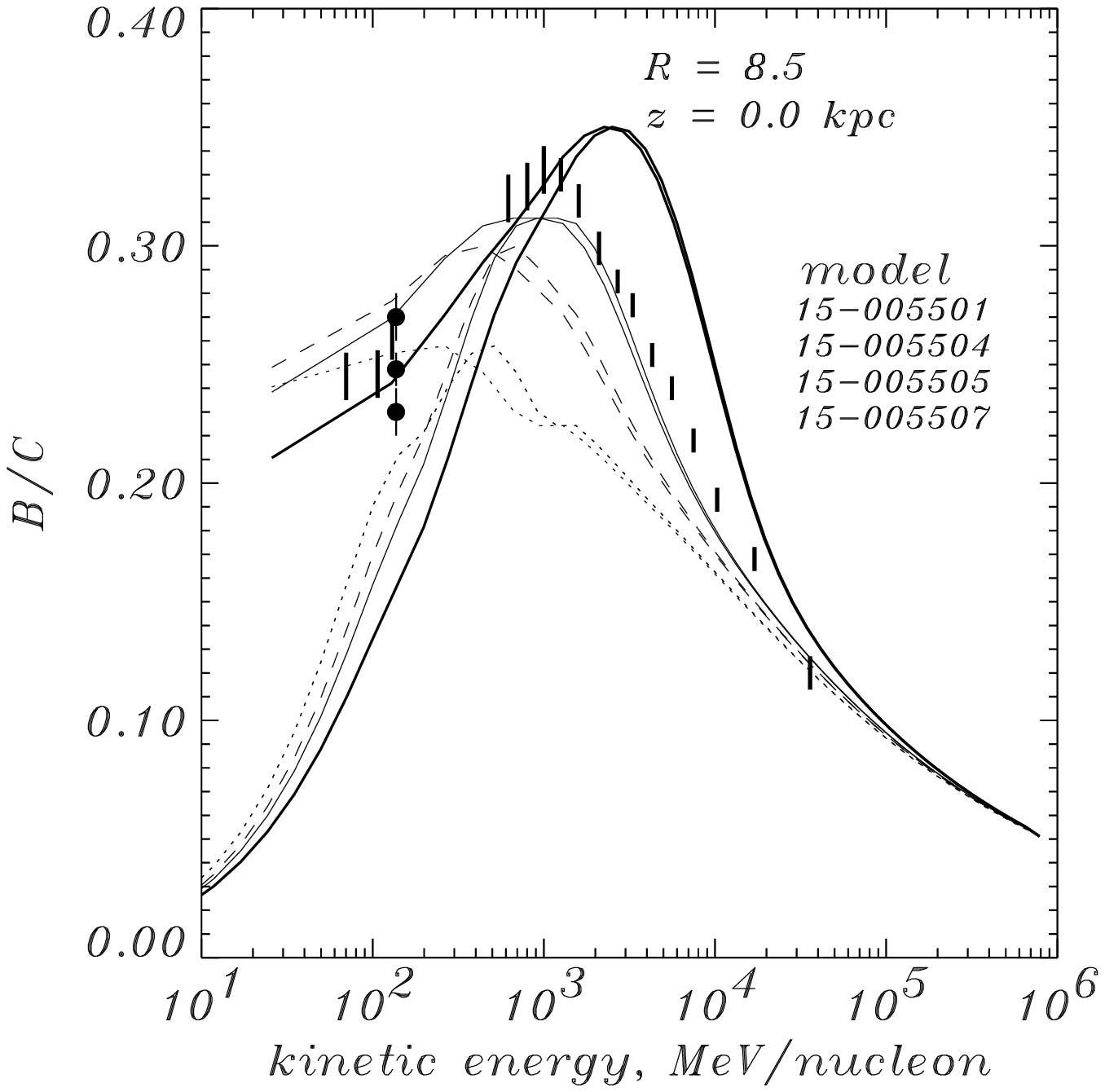}\hskip 2cm
\includegraphics[width=\fwc]{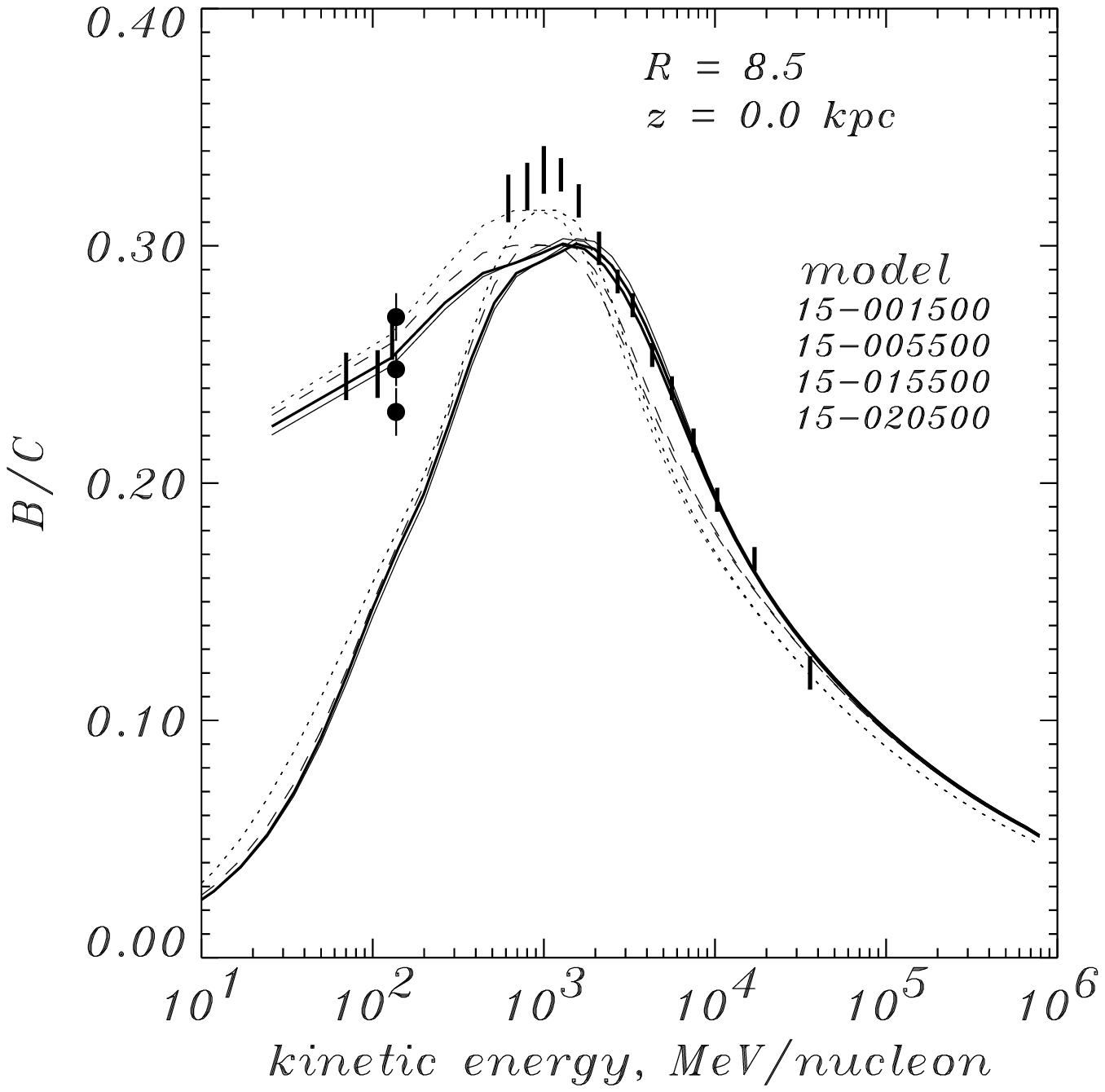}
}
\vskip -5pt
\parbox{89mm}{
\caption[fig7.ps]{  \footnotesize
\BC\ ratio for diffusive reacceleration models
with $z_h$ = 5 kpc, $v_A$ = 0 (dotted), 15 (dashed), 20 (thin solid), 30 km
s$^{-1}$ (thick solid).  Parameters as in Table~\ref{table2}.  In each
case the interstellar ratio and the ratio modulated to 500 MV is shown.
Data: as Figure~\ref{fig3}. \label{fig7} }
}\hspace{7mm}
\parbox{89mm}{
\caption[fig8.ps]{  \footnotesize
\BC\ ratio for diffusive reacceleration models:
$z_h = 1$ (dotted), 5 (dashed), 10 (thin solid), and 20 kpc (thick solid).
Parameters as in Table~\ref{table2}.  In each case the interstellar
ratio and the ratio modulated to 500 MV is shown.  Data: as
Figure~\ref{fig3}.\\  \label{fig8} }}
\end{figure*} 

\begin{figure*}[bht]
\centerline{
   \includegraphics[width=\fwb]{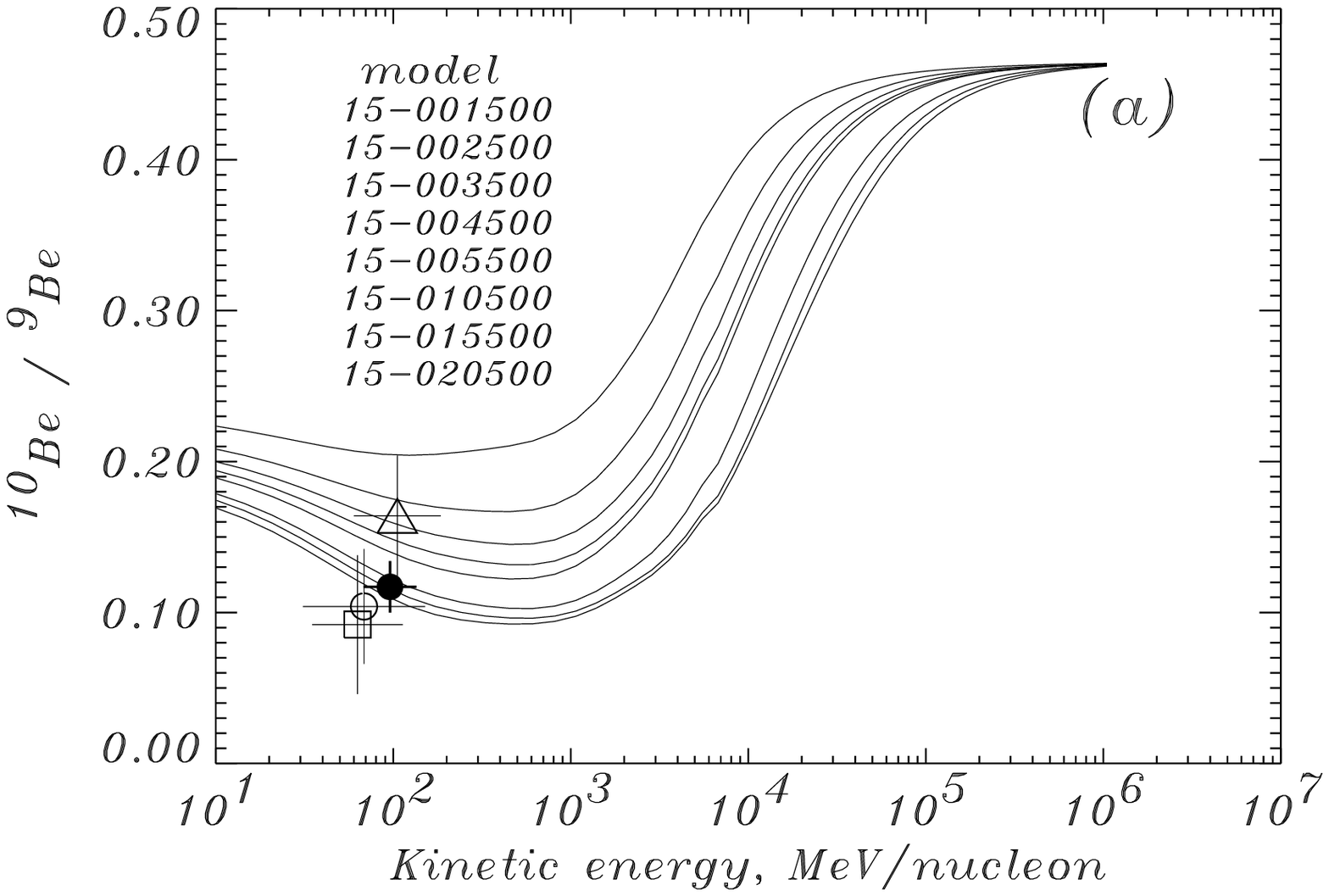}\hskip 2cm
   \includegraphics[width=\fwc]{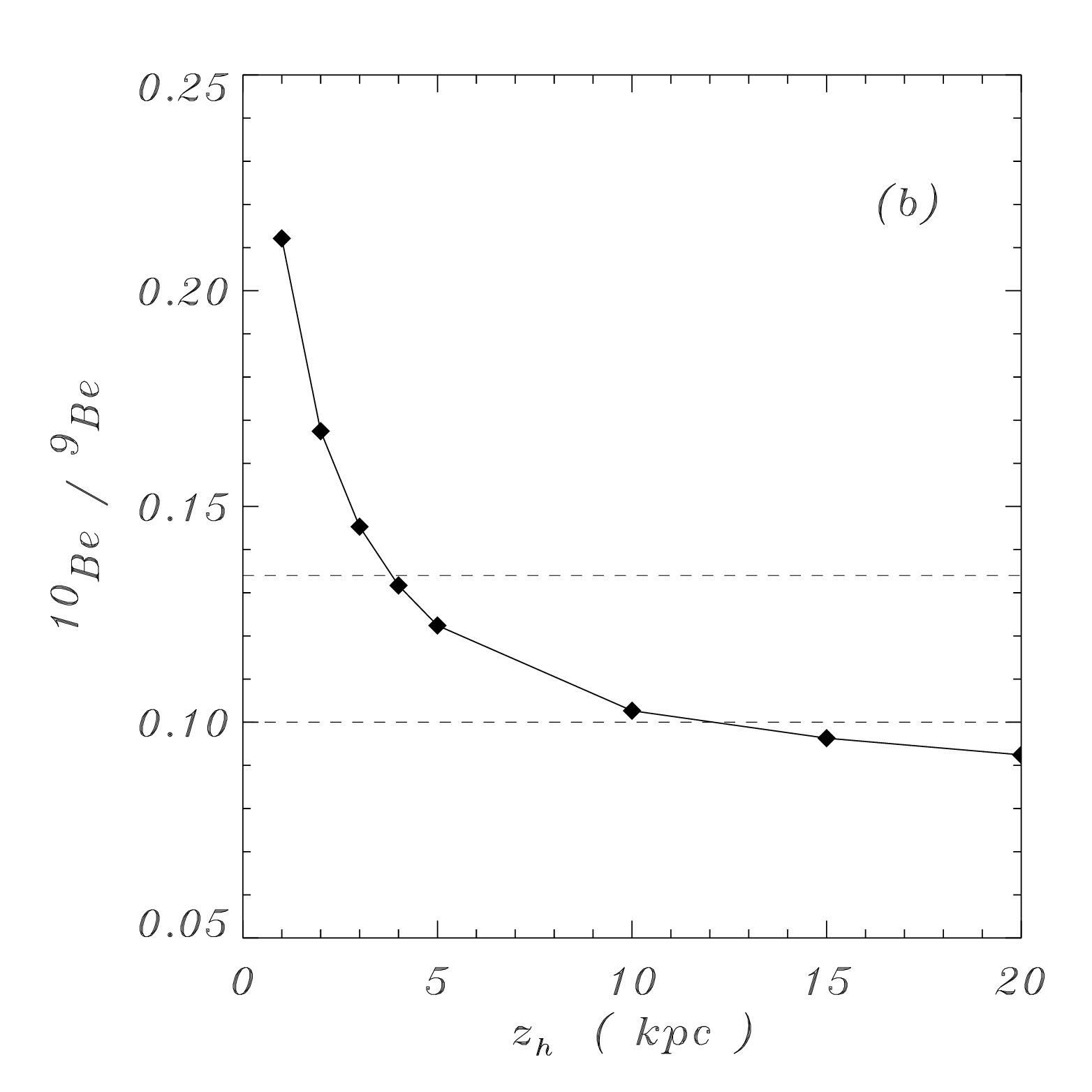}
}
\vskip -5pt
\caption[fig9a.ps,fig9b.ps]{ \footnotesize
\Berat\ ratio for diffusive
reacceleration models: (a) as function of energy for (from top to
bottom) $z_h$ = 1, 2, 3, 4, 5, 10, 15 and 20 kpc, data points as in
Figure~\ref{fig5}; (b) as function of $z_h$ at 525 MeV/nucleon
corresponding to the mean interstellar value for the Ulysses data
(\cite{Connell98}); the Ulysses experimental limits are shown as
horizontal dashed lines.  Parameters as in Table~\ref{table2}.
\label{fig9} }
\end{figure*}

Figure~\ref{fig5} shows the predicted and measured \Berat\ ratio; here
we use the models with a break in $\Dxx(p)$ since these do have the
correct \BC\ ratio in the few 100 MeV/nucleon range where the $Be$
measurements are available and are therefore appropriate for this
comparison independently of the situation at higher energies.  For our
final evaluation we use \Berat\ data from Ulysses, which has the
highest statistics.

Figure~\ref{fig6} summarizes the limits on $z_h$ and $dV/dz$, using the
\Berat\ ratio at the interstellar energy of 525 MeV/nucleon appropriate
to the Ulysses data (\cite{Connell98}).  For $z_h <4$ kpc, the
predicted ratio is always too high, even for no convection; no
convection is allowed for such $z_h$ values since this increases
\Berat\ still further.  For $z_h \ge 4$ kpc agreement with \Berat\ is
possible provided $0 < dV/dz < 7$ km s$^{-1}$ kpc$^{-1}$.  We conclude
from Figure~\ref{fig6}a that in the absence of convection
$4{\rm\ kpc}<z_h < 12 {\rm\ kpc}$, and if convection is allowed the
lower limit remains but no upper limit can be set.  It is interesting
that an upper as well as a lower limit on $z_h$ is obtained in the case
of no convection, although \Berat\ approaches asymptotically a constant
value for large halo sizes and becomes insensitive to the halo
dimension. From Figure~\ref{fig6}b, $dV/dz < 7$ km s$^{-1}$ kpc$^{-1}$
and this figure places upper limits on the convection parameter for
each halo size. These limits are rather strict, and a finite wind
velocity is only allowed in any case for $z_h > 4$ kpc.  Note that
these results are not very sensitive to modulation since the predicted
\Berat\ is fairly constant from 100 to 1000 MeV/nucleon.

Our results can be compared with those of other studies: $z_h \ge7.8$
kpc (\cite{Freedman80}), $ z_h \le3$ kpc (\cite{Bloemen93}), and $z_h
\le4$ kpc (\cite{Webber92}). Most recently Lukasiak et al. (1994a)
found $1.9{\rm\ kpc} < z_h < 3.6$ kpc (for no convection) based on
Voyager $Be$ data and using the Webber, Lee \& Gupta (1992) models.  We
believe our new limits to be an improvement, first because of the
improved $Be$ data from Ulysses, second because of our treatment of
energy losses (see Section~\ref{diff_reacc_models}) and generally more
realistic astrophysical details in our model.  The papers cited also
did not consider the low-energy \BC\ data, which we have shown are in
fact a problem for diffusion/convection models.

The cosmic-ray driven wind models of Zirakashvili et al. (1996) have
values of $dV/dz \approx 10$ km s$^{-1}$ kpc$^{-1}$, somewhat larger
than our upper limits.  Since their models are different from ours in
many respects this is not significant, but suggests it would be useful
to carry out calculations like those in the present paper for such
models to provide a critical test of their viability.

\begin{deluxetable}{ccccccc}
\tablefontsize{\footnotesize}
\tablecolumns{7}
\footnotesize

\tablecaption{Parameters of diffusion/convection models.  \label{table1}}

\tablehead{
\colhead{Model} & \colhead{$z_h$} & \colhead{$D_0$} & \colhead{$\rho_0$}
 & \colhead{$\delta_1$} & \colhead{$\delta_2$} & \colhead{$dV/dz$} \\
\colhead{} & \colhead{kpc} & \colhead{$10^{28}$ cm$^2$ s$^{-1}$} & \colhead{GV}
 & \colhead{} & \colhead{} & \colhead{km s$^{-1}$ kpc$^{-1}$}
}
\startdata

%
%

01000   & 1 &0.7 &  3  &  0.60 &  0.60 &  0   \nl
01010   & 1 &0.7 &  3  &  0.60 &  0.60 & 10   \nl
01020   & 1 &0.7 &  3  &  0.60 &  0.60 & 20   \nl
\noalign{\medskip}
03000   & 3 &2.0 &  3  &  0.60 &  0.60 &  0   \nl
03010   & 3 &1.4 &  3  &  0.65 &  0.65 & 10   \nl
03020   & 3 &1.1 &  3  &  0.70 &  0.70 & 20   \nl
\noalign{\medskip}
10000   &10 &5.0 &  3  &  0.60 &  0.60 &  0   \nl
10010   &10 &2.5 &  3  &  0.70 &  0.70 & 10   \nl
10020   &10 &1.1 &  3  &  0.90 &  0.90 & 20   \nl
\noalign{\medskip}
01100   & 1 &0.9 &  5  & --0.60&  0.60 &  0   \nl   
01105   & 1 &0.8 &  5  & --0.60&  0.60 &  5   \nl
01110   & 1 &0.8 &  5  & --0.60&  0.60 & 10   \nl
\noalign{\medskip}
03100   & 3 &2.5 &  5  & --0.60&  0.60 &  0   \nl  
03105   & 3 &2.2 &  5  & --0.60&  0.60 &  5   \nl
03110   & 3 &2.0 &  5  & --0.60&  0.60 & 10   \nl
\noalign{\medskip}
04100   & 4 &3.5 &  5  & --0.60&  0.60 &  0   \nl  
04105   & 4 &2.7 &  5  & --0.60&  0.70 &  5   \nl
04110   & 4 &2.5 &  5  & --0.60&  0.70 & 10   \nl
\noalign{\medskip}
05100   & 5 &4.5 &  5  & --0.60&  0.60 &  0   \nl  
05105   & 5 &3.2 &  5  & --0.60&  0.70 &  5   \nl
05110   & 5 &2.5 &  5  & --0.60&  0.70 & 10   \nl
\noalign{\medskip}
10100   &10 &7.0 &  5  & --0.60&  0.60 &  0   \nl
10105   &10 &3.8 &  5  & --0.60&  0.80 &  5   \nl
10110   &10 &3.0 &  5  & --0.60&  0.80 & 10   \nl
\noalign{\medskip}
15100   &15 &9.0 &  5  & --0.60&  0.60 &  0   \nl
15105   &15 &3.8 &  5  & --0.60&  0.80 &  5   \nl
15110   &15 &3.0 &  5  & --0.60&  0.80 & 10   \nl
\noalign{\medskip}
20100   &20 &9.0 &  5  & --0.60&  0.60 &  0   \nl
20105   &20 &3.8 &  5  & --0.60&  0.80 &  5   \nl
20110   &20 &3.0 &  5  & --0.60&  0.80 & 10   \nl

\enddata
\end{deluxetable}

\begin{deluxetable}{cccccc}
\tablefontsize{\footnotesize}
\tablecolumns{6}
\footnotesize

\tablecaption{ Parameters of diffusive reacceleration models\tablenotemark{a}.
    \label{table2}}

\tablehead{
\colhead{Best fit} & \colhead{Models with} & \colhead{Models with SNR} 
 & \colhead{$z_h$} & \colhead{$D_0$} & \colhead{$v_A$} \\
\colhead{models\tablenotemark{b}} & \colhead{no energy losses\tablenotemark{b}} 
 & \colhead{source distribution\tablenotemark{c}}
 & \colhead{kpc} & \colhead{$10^{28}$ cm$^2$ s$^{-1}$} & \colhead{km s$^{-1}$}
}
\startdata

01500   & 01510   & 01511   & 1 &  1.7 & 20 \nl
02500   & 02510   & 02511   & 2 &  3.2 & 20 \nl
03500   & 03510   & 03511   & 3 &  4.6 & 20 \nl
04500   & 04510   & 04511   & 4 &  6.0 & 20 \nl 
05500   & 05510   & 05511   & 5 &  7.7 & 20 \nl
10500   & 10510   & 10511   &10 &  12  & 20 \nl 
15500   & 15510   & 15511   &15 &  15  & 20 \nl 
20500   & 20510   & 20511   &20 &  16  & 18 \nl 
\noalign{\medskip}
\multicolumn{3}{l}{Effect of varying $v_A$:}\nl
05501   & \nodata & \nodata & 5 &  7.7 &  0 \nl 
05502   & \nodata & \nodata & 5 &  7.7 &  5 \nl 
05503   & \nodata & \nodata & 5 &  7.7 & 10 \nl 
05504   & \nodata & \nodata & 5 &  7.7 & 15 \nl 
05505   & \nodata & \nodata & 5 &  7.7 & 20 \nl 
05506   & \nodata & \nodata & 5 &  7.7 & 25 \nl 
05507   & \nodata & \nodata & 5 &  7.7 & 30 \nl 
\noalign{\medskip}

\enddata

\footnotesize
\tablenotetext{a}{For all reacceleration models $\rho_0=3$ GV, $\delta=1/3$ 
   (see Section~2 for details)}
\tablenotetext{b}{Parameters of the source distribution (eq.~[\ref{2.2}]):
   $\eta=0.5, \xi=1.0$}
\tablenotetext{c}{Parameters of the SNR distribution (eq.~[\ref{2.2}]):
   $\eta=1.69, \xi=3.33$}

\end{deluxetable}

\subsection{Diffusive reacceleration models \label{diff_reacc_models}}

The main parameters are $z_h$, $D_0$ and $v_A$ ($\rho_0$ is arbitrary
since $\delta$ is constant).  Again we treat $z_h$ as the main unknown
quantity.  The evaluation is simpler than for convection models since
the number of free parameters is smaller.
The parameters of these models are summarized in Table~\ref{table2}.
Figure~\ref{fig7} illustrates the effect on \BC\ of varying $v_A$, from
$v_A = 0$ (no reacceleration) to $v_A=30$ km s$^{-1}$, for $z_h= 5$
kpc.  This shows how the initial form becomes modified to produce
the characteristic peaked shape.  Reacceleration models thus lead
naturally to the observed peaked form of \BC, as pointed out by several
previous authors (e.g., \cite{Letaw93,SeoPtuskin94,HeinbachSimon95}).

Figure~\ref{fig8} shows \BC\ for $z_h = 1 - 20$ kpc.  Our value of $v_A
\approx 20$ km~s$^{-1}$ is consistent with the value obtained by Seo \&
Ptuskin (1994) which they also derived from \BC; since for stable
nuclei the leaky-box and diffusion treatments are equivalent this is a
good test of the operation of our code.  The value of $v_A$ is typical
of the warm ionized phase of the interstellar gas
(\cite{SeoPtuskin94}).  The exact low-energy form of \BC\ depends on
details of the modulation so that an exact fit here is not attempted;
note however that $v_A$ and $D_0$ can be (and indeed were) determined
from the high-energy \BC\ alone, and the low-energy agreement is then
satisfactory\footnote{Since we are considering a {\it ratio} at the
same rigidity the effect of modulation is confined to a deceleration
$\approx$ 200 MeV/nucleon (cf. spectra where absolute intensity changes
are important).}.  Figure~\ref{fig9} shows \Berat\ for the same models,
(a) as a function of energy for various $z_h$, (b) as a function of
$z_h$ at 525 MeV/nucleon corresponding to the Ulysses measurement.
Comparing with the Ulysses data point, we conclude that $4{\rm\ kpc}
<z_h < 12$ kpc. Again the result is not very sensitive to modulation
since the predicted \Berat\ is fairly constant from 100 to 1000
MeV/nucleon.

\medskip
\centerline{\includegraphics[width=\fwc]{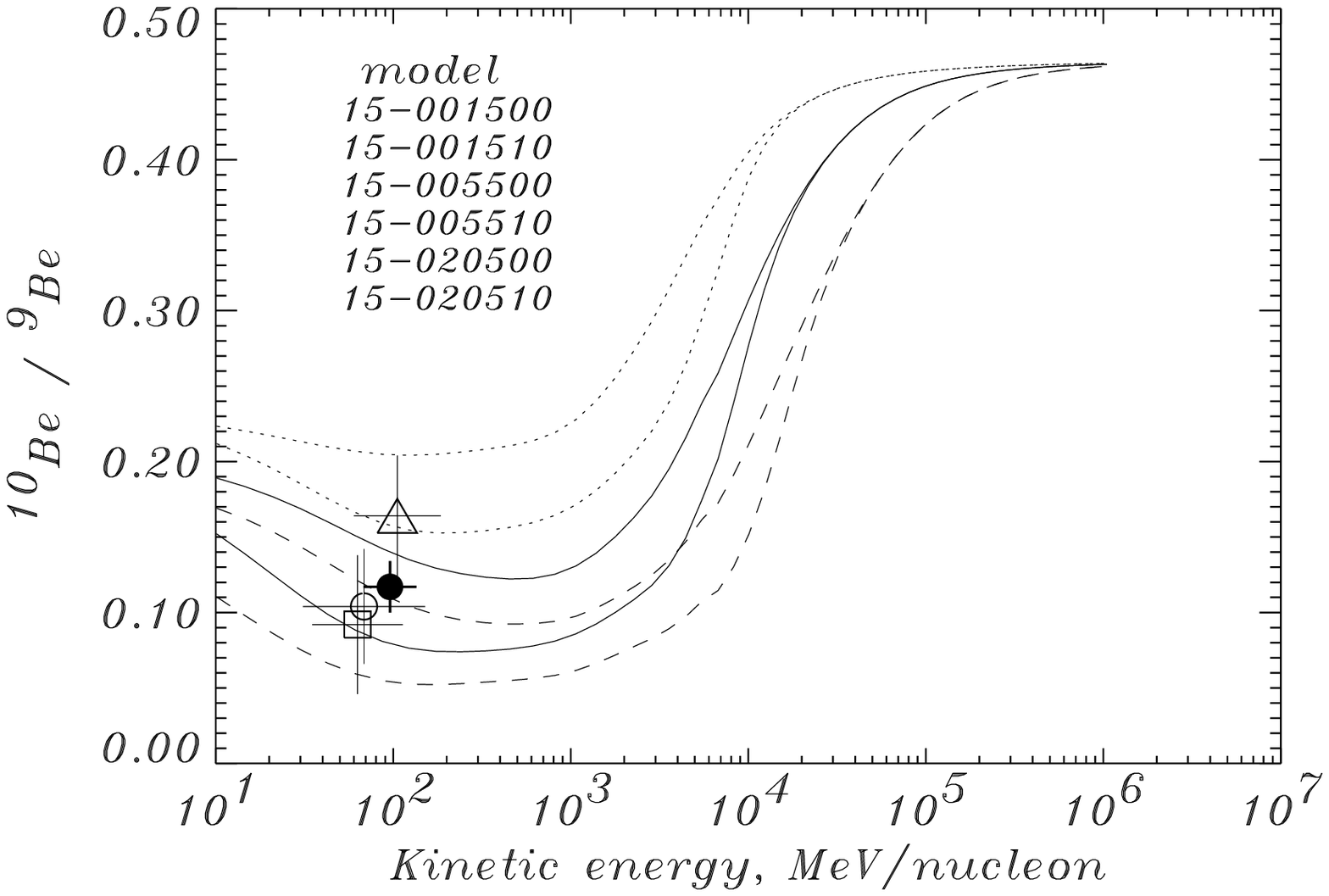}}
\vskip -5pt
\figcaption[fig10.ps]{ \Berat\ ratio
for diffusive reacceleration model, showing influence of energy losses,
for $z_h=1$ kpc (dotted lines), 5 kpc (solid), 20 kpc (dashed).  In
each case upper curve is with energy losses, lower curve without.
Parameters as in Table~\ref{table2}.  Data points as in
Figure~\ref{fig5}.  \label{fig10} }
\medskip

Figure~\ref{fig10} illustrates the importance of energy losses on the
\Berat\ ratio; for reacceleration cases with $z_h = 1 - 20$ kpc, we
show the ratio with and without losses. Losses attenuate the flux of
stable nuclei much more than radioactive nuclei, and hence lead to an
increase in \Berat.  The effect can be simply illustrated as follows.
The ionization loss rate on neutral gas is $\sim 1.8\times 10^{-7}Z^2
\left< n_H\right>\beta^{-1}$ eV s$^{-1}$, where $\beta=v/c$ is the
nucleon speed, and $\left< n_H\right>$ is the average interstellar gas
density.  Thus for $Be$-nuclei of 300 MeV/nucleon and for a gas disk
with 0.2 kpc thickness and density 1 cm$^{-3}$, $\left<
n_H\right>=0.2/z_h$ cm$^{-3}$, which gives the loss time of $\sim 3
\times 10^8$ years for $z_h = 5$ kpc.  Coulomb losses on the ionized
gas in the halo increase the losses further (see Figure~\ref{fig13});
although the density is low the wide $z$-extent means that the losses
occur over large regions of the halo.  For the same $z_h$ the
diffusion time is $\approx 4 \times 10^8$ years so the stable $^9Be$ is
significantly attenuated.  For the radioactive $^{10}Be$ ($\tau_{1/2} =
1.6 \times 10^6$ years) the energy losses are negligible. Hence losses
significantly increase \Berat.  As can be seen in Figure~\ref{fig10},
the relative effect is largest for large halos and becomes a dominant
effect only for $z_h > 3$ kpc.  Although we illustrate this for the
reacceleration case, the same effect applies to diffusion/convection
models.  Clearly if losses are ignored the predicted ratio will be too
low and the derived value of $z_h$ will be too small since $z_h$ will have
to be reduced to fit the observations.

The proton, Helium and positron spectra were presented in
\cite{MoskalenkoStrong98a} for the case $z_h= 3$ kpc using the same
model as used here, and the injection spectra were derived.  The effect
of varying the halo size is small for these spectra so we do not extend
that calculation to different $z_h$.

\section{Cosmic-ray gradients \label{CRgradients}}
An important constraint on any model of cosmic-ray propagation is
provided by gamma-ray data which give information on the radial
distribution of cosmic rays in the Galaxy. For a given source
distribution, a large halo will give a smaller cosmic-ray gradient.  It
is generally believed that supernova remnants (SNR) are the main
sources of cosmic rays (see \cite{Webber97} for a recent review), but
unfortunately the distribution of SNR is poorly known due to selection
effects.  Nevertheless it is interesting to compare quantitatively the
effects of halo size on the gradient for a plausible SNR source
distribution.  For illustration we use the SNR distribution from Case
\& Bhattacharya (1996), which is peaked at $R = 4 - 5$ kpc and has a
steep falloff towards larger $R$.

Figure~\ref{fig11} shows the effect of halo size on the resulting
radial distribution of 3 GeV cosmic-ray protons, for the reacceleration
model.  For comparison we show the cosmic-ray distribution deduced by
model-fitting to EGRET gamma-ray data ($>100$ MeV) from Strong \&
Mattox (1996), which is dominated by the $\pi^o$-decay component
generated by GeV nucleons; the analysis by Hunter et al. (1997), based
on a different approach, gives a similar result.  The predicted
cosmic-ray distribution using the SNR source function is too steep even
for large halo sizes; in fact the halo size has a relatively small
effect on the distribution.  Other related distributions such as pulsars
(\cite{Taylor93,Johnston94}) have an even steeper falloff.  Only for
$z_h = 20$ kpc does the gradient approach that observed, and in this
case the combination of a large halo and a slightly less steep SNR
distribution could give a satisfactory fit.  For diffusion/convection
models the situation is similar, with more convection tending to make
the gradient follow more closely the sources.
A larger halo ($z_h \gg 20$ kpc), apart from being excluded by the
$^{10}Be$ analysis presented here, would in fact not improve the
situation much since Fig.~\ref{fig11} shows that the gradient
approaches an asymptotic shape which hardly changes beyond a certain
halo size.  This is a consequence of the nature of the diffusive
process, which even for an unlimited propagation region still retains
the signature of the source distribution.

\medskip
\centerline{\includegraphics[width=\fwc]{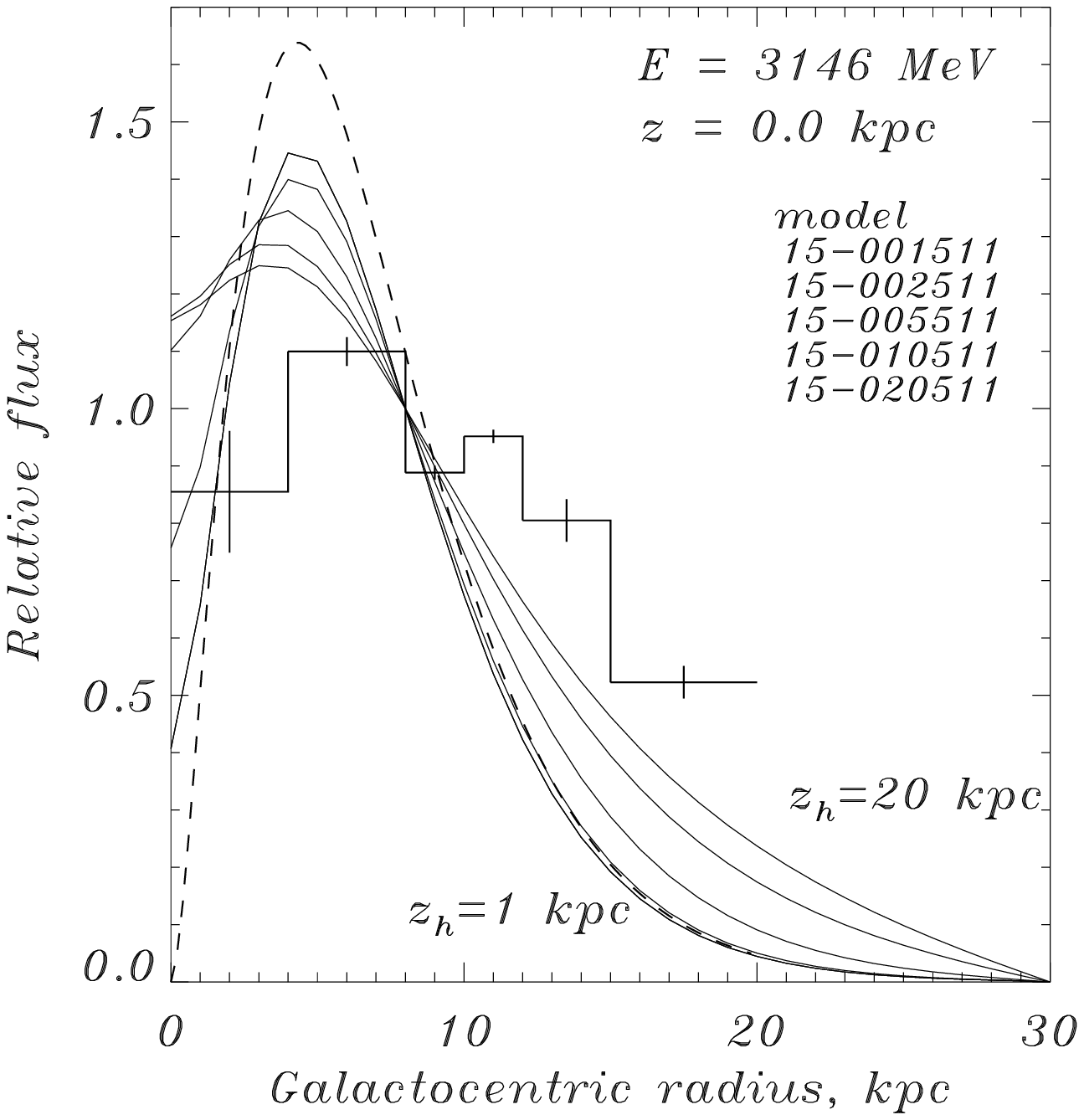}}
\vskip -5pt
\figcaption[fig11.ps]{ Radial distribution of 3 GeV protons at $z = 0$,
for diffusive reacceleration model with halo sizes $z_h = 1$, 3, 5,
10, 15, and 20 kpc (solid curves). The source distribution is that for
SNR given by Case \& Bhattacharya (1996), shown as a dashed line. The
cosmic-ray distribution deduced from EGRET $>$100 MeV gamma rays
(\cite{StrongMattox96}) is shown as the histogram.  Parameters as in
Table~\ref{table2}.  \label{fig11} }
\medskip

Based on these results we have to conclude, in the context of the
present models, that the distribution of sources is not that
expected from the (highly uncertain: see \cite{Green91}) distribution of SNR. 
This conclusion is similar to that previously
found by others (\cite{Webber92,Bloemen93}).  In view of the difficulty
of deriving the SNR distribution this is perhaps not a serious
shortcoming; if SNR are indeed CR sources then it is possible that the
gamma-ray analysis gives the best estimate of their Galactic
distribution.  Therefore in our standard model we have obtained the
source distribution empirically by requiring consistency with the
high-energy gamma-ray results.

Figure~\ref{fig12} shows the source distribution adopted in the present
work, and the resulting 3 GeV proton distribution, again compared to
that deduced from gamma rays. The gradients are now consistent, 
especially considering that some systematic effects, due for example
unresolved gamma-ray sources, are present in the gamma-ray results.

Measurements of cosmic-ray anisotropy in the 1 -- 100 TeV range provide
an independent argument for reacceleration (e.g., \cite{SeoPtuskin94})
since the slower increase of diffusion coefficient with energy avoids
the large anisotropies predicted by non-reacceleration models.  Our
models reproduce this behaviour, the reacceleration models giving
anisotropies $\sim 10^{-3}$ at 1 TeV, while the non-reacceleration
models give $> 10^{-2}$.  The observed values ($\sim 10^{-3}$)
largely reflect the local structure of the interstellar magnetic field
in the part of the Galaxy near the Sun, and hence do not give useful
constraints on the large-scale propagation which our model addresses
(see \cite{Berezinskii90}).  In particular it is not possible to
test the large-scale cosmic-ray gradients at such energies by this
method.  It is sufficient to note that the reacceleration models are
consistent with the observations while the non-reacceleration models
are not, in accord with previous authors' conclusions.

\medskip
\centerline{\includegraphics[width=\fwc]{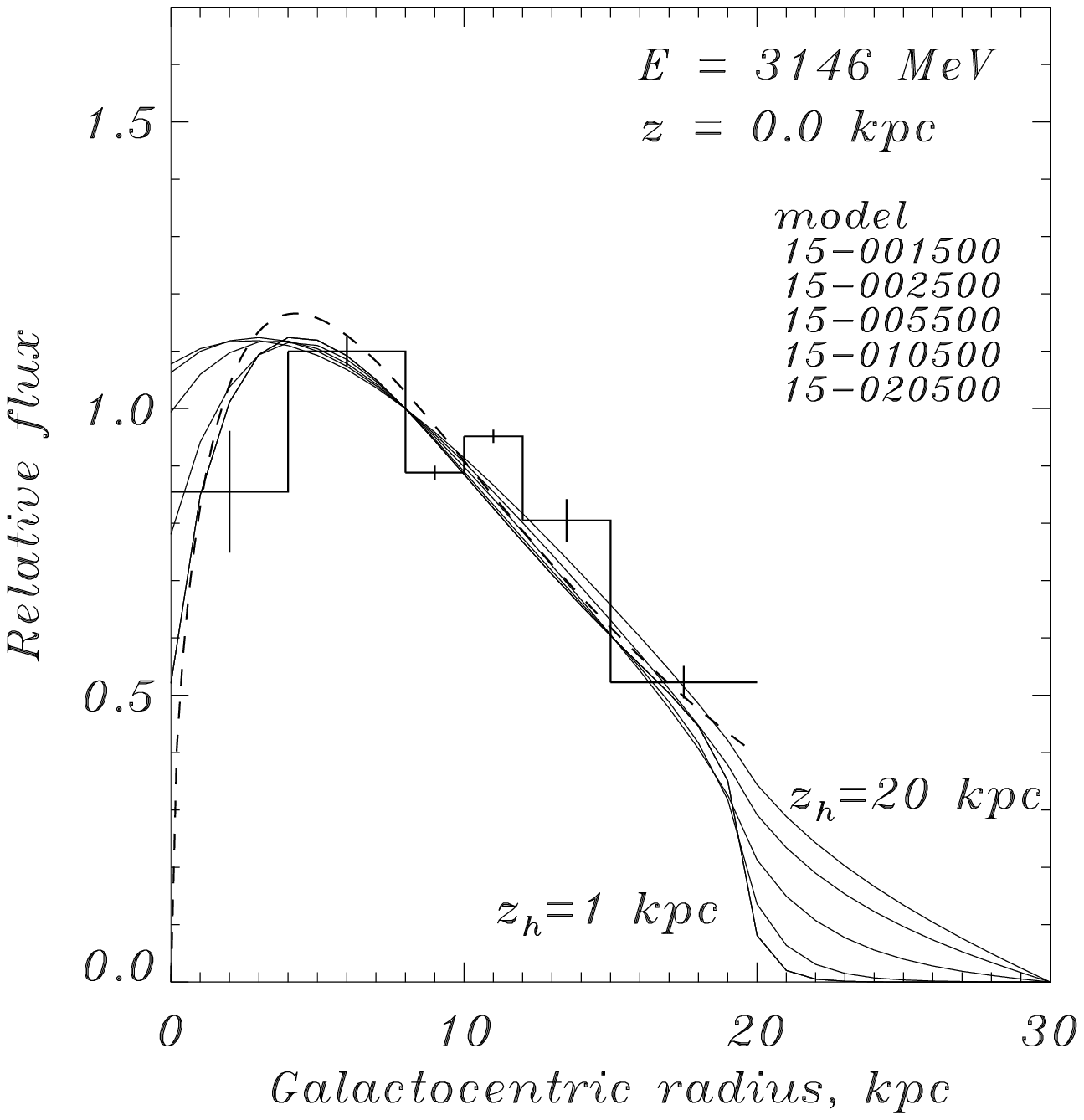}}
\vskip -5pt
\figcaption[fig12.ps]{ Radial distribution of 3 GeV protons at $z = 0$,
for diffusive reacceleration model with various halo sizes $z_h = 1$,
3, 5, 10, 15, and 20 kpc (solid curves). The source distribution used
is shown as a dashed line, and is that adopted to reproduce the
cosmic-ray distribution deduced from EGRET $>$100 MeV gamma rays
(\cite{StrongMattox96}), shown as the histogram.  Parameters as in
Table~\ref{table2}.  \label{fig12} }
\medskip

\section{Conclusions}
We have shown that simple diffusion/convection models have difficulty
in accounting for the observed form of the \BC\ ratio without special
assumptions chosen to fit the data, and do not obviate the need for an
{\it ad hoc} form for the diffusion coefficient.  On the other hand we
confirm the conclusion of other authors that models with reacceleration
account naturally for the energy dependence over the whole observed
range, with only two free parameters.  Combining these results points
rather strongly in favour of the reacceleration picture. In this case
$v_A \approx 20 $ km s$^{-1}$, with little dependence on $z_h$.

For the first time \Berat\ has also been computed with reacceleration.
We take advantage of the recent Ulysses $Be$ measurements to improve
limits on the halo size.  We emphasize the crucial importance of the
treatment of energy losses in the evaluation of the \Berat\ ratio.  The
halo height with reacceleration is $4{\rm\ kpc} < z_h < 12$ kpc.  Our
new limits should be an improvement on previous estimates because of
the more accurate $Be$ data, our treatment of energy losses, and the
inclusion of more realistic astrophysical details (such as, e.g., the
gas distribution) in our model.

In case reacceleration is {\it not} important, the halo size limits are
still $4{\rm\ kpc} < z_h < 12$ kpc for the case of no convection, while
only the lower limit holds if convection is allowed.  The upper limit
on the convection velocity gradient is $dV/dz < 7$ km s$^{-1}$
kpc$^{-1}$, and this value being allowed for large $z_h$ only.

The gradient of protons derived from gamma rays is smaller than
expected for SNR sources, the closest approach to consistency being
for $z_h = 20$ kpc; we therefore adopt a flatter source distribution
in order to meet the gamma-ray constraints.

The anisotropy at $\sim 1$ TeV predicted by our reacceleration models
is consistent with observations, while the non-reacceleration model
predict a larger value than observed.  This reflects the general
property of such models (e.g., \cite{SeoPtuskin94}).  The large-scale
propagation is however not significantly constrained by anisotropy
measurements at the energies considered in this paper, since local
interstellar effects may dominate.

The large $z_h$ values found here would have very significant
implications for gamma rays at high galactic latitudes, giving a larger
inverse-Compton intensity than normally considered. Gamma-rays will be
addressed in detail in \cite{MoskalenkoStrong98b}.

\acknowledgements
We are grateful to the referee for useful suggestions.  We thank Dr. J.
J. Connell for help with the Ulysses $Be$ data and for providing data
prior to publication. We thank Dr. D. Breitschwerdt and Dr. V. Ptuskin
for useful discussions.

\onecolumn
\appendix
\section*{A. Propagation equation \label{propagation_eq}}

The propagation equation is written in the form:
\begin{equation}
\label{A.1}
{\partial \psi \over \partial t} 
= q(\vec r, p) 
+ \vec\nabla \cdot ( \Dxx\vec\nabla\psi - \vec V\psi )
+ \ddp\, p^2 \Dpp \ddp\, {1\over p^2}\, \psi
- {\partial\over\partial p} \left[\dot{p} \psi
- {p\over 3} \, (\vec\nabla \cdot \vec V )\psi\right]
- {1\over\tau_f}\psi - {1\over\tau_r}\psi\ ,
\end{equation}
where $\psi=\psi (\vec r,p,t)$ is the density per unit of total
particle momentum, $\psi(p)dp = 4\pi p^2 f(\vec p)$ in terms of
phase-space density $f(\vec p)$, $q(\vec r, p)$ is the source term,
$\Dxx$ is the spatial diffusion coefficient, $\vec V$ is the convection
velocity, reacceleration is described as diffusion in momentum space
and is determined by the coefficient $\Dpp$, $\dot{p}\equiv dp/dt$
is the momentum loss rate, $\tau_f$ is the time scale for
fragmentation, and $\tau_r$ is the time scale for the radioactive
decay.  The details of the numerical scheme is described in 
Appendix~B.

We use particle momentum as the kinematic variable since it greatly
facilitates the inclusion of the diffusive reacceleration terms.  The
injection spectrum of primary nucleons is assumed to be a power law in
momentum for the different species, $dq(p)/dp \propto p^{-\Gamma}$ for
the injected {\it density}\footnote{ This corresponds to an injected
{\it flux} $dF(p)/dp \propto \beta p^{-\Gamma}$ or $dF(E_k)/dE_k
\propto p^{-\Gamma}$, a form often used (e.g., Engelmann et al.
1990).  Since observations are usually quoted as a {\it flux} with
kinetic energy per nucleon as the kinematic variable a conversion is
made before comparison with data: $dF(E_k)/dE_k = {c\over4\pi} \beta
\psi (dp/dE_k) = {c\over4\pi} A \psi$ since $dp/dE_k = A/\beta$, where
A is the mass number, $E_k$ is the kinetic energy per nucleon,
$\beta=v/c$.}, as expected for diffusive shock acceleration (e.g.,
\cite{Blandford80}); the value of $\Gamma$ can vary with species.  The
injection spectrum for $^{12}C$ and $^{16}O$ was taken as $dq(p)/dp
\propto p^{-2.35}$, for the case of no reacceleration, and $p^{-2.25}$
with reacceleration.  These values are consistent with Engelmann et
al.\ (1990) who give an injection index $2.23\pm0.05$. The same indices
reproduce the observed proton and $^4He$ spectra, as was shown in
\cite{MoskalenkoStrong98a}.  For primary electrons, the injection
spectrum is adjusted to reproduce direct measurements, gamma-ray and
synchrotron data; details are given in the other papers of this series
(I, II, IV).

For secondary nucleons, the source term is $q(\vec r, p) = \beta c\,
\psi_p (\vec r, p)[\sigma^{ps}_H (p) n_H (\vec r)+ \sigma^{ps}_{He}(p)
n_{He}(\vec r)]$, where $\sigma^{ps}_H (p)$, $\sigma^{ps}_{He} (p)$ are
the production cross sections for the secondary from the progenitor on
H and He targets, $\psi_p$ is the progenitor density, and $n_H$,
$n_{He}$ are the interstellar hydrogen and Helium number densities.

To compute \BC\ and \Berat\ it is sufficient for our purposes to treat
only one principal progenitor and compute weighted cross sections based
on the observed cosmic-ray abundances, which we took from Lukasiak et
al. (1994b).  Explicitly, for a principal primary with abundance $I_p$,
we use for the production cross section $\overline\sigma^{ps} = \sum_i
\sigma^{is} I_i/I_p$, where $\sigma^{is}$, $I_i$ are the cross sections
and abundances of all species producing the given secondary.  For the
case of Boron, the Nitrogen progenitor is secondary but only accounts
for $\approx$ 10\% of the total Boron production, so that the
approximation of weighted cross sections is sufficient.

For the fragmentation cross sections we use the formula given by Letaw,
Silberberg \& Tsao (1983).  For the secondary production cross sections
we use the Webber, Kish \& Schrier (1990) and Silberberg \& Tsao (1990,
see also \cite{Garcia87}) parameterizations in the the form of code
obtained from the Transport Collaboration (\cite{Guzik97}).  Comparison
of the results from these different versions of the cross sections
gives a useful estimate of the uncertainty from this source.  For the
important \BC\ ratio, we take the $^{12}C$, $^{16}O \to\, ^{10}B$,
$^{10}C$, $^{11}B$, $^{11}C$ cross sections from the fit to
experimental data given by Heinbach \& Simon (1995).  Since for $Be$
the values of the cross sections are particularly important we give for
reference the values actually used for the abundance-weighted cross
sections at 500 MeV/nucleon, including interstellar $He$:
$\overline\sigma(^{12}C\to\,^{9}Be) = 18.2$ mb,
$\overline\sigma(^{12}C\to\,^{10}Be)= 8.6$ mb.  For radioactive decay,
$\tau_r = \gamma\tau_{1/2}/\ln2$, where $\tau_{1/2} = 1.6\times 10^6$
years for $^{10}Be$.

For electrons and positrons the same propagation equation is valid when
the appropriate energy loss terms (ionization, bremsstrahlung, inverse
Compton, synchrotron) are used. Since this paper is intended to
complete the description of the full model, we include the formulae for
these loss mechanisms in Appendix~C.2. A
detailed description of the source function for secondary electrons and
positrons was given in \cite{MoskalenkoStrong98a}.

\section*{B. Numerical solution of propagation equation \label{numerical_solution}}
The diffusion, reacceleration, convection and loss terms in
eq.~(A.1) can all be finite-differenced for each dimension ($R, z,
p$) in the form
\begin{equation}
\label{B.1}
{\partial \psi_i\over\partial t}
= {\psi^{t+\Delta t}_i -\psi^t_i \over \Delta t}
= {\alpha_1\psi^{t+\Delta t}_{i-1} -\alpha_2\psi^{t+\Delta t}_i 
+ \alpha_3\psi^{t+\Delta t}_{i+1}\over\Delta t}+ q_i \ ,
\end{equation}
where all terms are functions of $(R, z, p)$.

In the Crank-Nicholson implicit method (\cite{Press92}) the updating
scheme is
\begin{equation}
\label{B.2}
\psi_i^{t+\Delta t} = \psi_i^t + \alpha_1 \psi_{i-1}^{t+\Delta t} - 
\alpha_2 \psi_i^{t+\Delta t} +\alpha_3 \psi_{i+1}^{t+\Delta t} + q_i
\Delta t \ .
\end{equation}
The tridiagonal system of equations,
\begin{equation}
\label{B.3}
- \alpha_1 \psi_{i-1}^{t+\Delta t} 
+ ( 1 + \alpha_2) \psi_i^{t+\Delta t}
- \alpha_3 \psi_{i+1}^{t+\Delta t} 
= \psi_i^t + q_i \Delta t ,
\end{equation}
is solved for the $\psi_i^{t+\Delta t}$ by the standard method
(\cite{Press92}).  Note that for energy losses we use `upwind'
differencing to enhance stability, which is possible since we have
only {\it loss} terms (adiabatic energy {\it gain} is not included
here).

The three spatial boundary conditions 
\begin{equation}
\label{B.4}
\psi(R,z_h,p) = \psi(R,-z_h,p) = \psi(R_h,z,p) = 0
\end{equation}
are imposed at each iteration. No boundary conditions are imposed or
required at $R$ = 0 or in $p$.  Grid intervals are typically $\Delta R
= 1$ kpc, $\Delta z = 0.1$ kpc; for $p$ a logarithmic scale with ratio
typically 1.2 is used.  Although the model is symmetric around $z = 0$
the solution is generated for $-z_h < z < z_h$ since this is required
for the tridiagonal system to be valid.

Since we have a 3-dimensional $(R,z,p)$ problem we use `operator
splitting' to handle the implicit solution, as follows.  We apply the
implicit updating scheme alternately for the operator in each dimension
in turn, keeping the other two coordinates fixed.  To account for the
substeps ${1\over 3} q_i$ and $1\over3\tau$ are used instead of $q_i$,
$1/\tau$.  The coefficients of the Crank-Nicholson scheme we use are given 
in Table~\ref{table3}.

The method was found to be stable for all $\alpha$, and this property
can be exploited to advantage by starting with $\alpha\gg 1$ (see
below).  The standard alternating direction implicit (ADI) method, in
which the full operator is used to update each dimension implicitly in
turn, is more accurate but was found to be unstable for $\alpha>1$.
This is a disadvantage when treating problems with many timescales, but
can be used to generate an accurate solution from an approximation
generated by the non-ADI method.

A check for convergence is performed by computing the timescale
${\psi\over \partial \psi/\partial t}$ from eq.~(A.1) and
requiring that this be large compared to all diffusive and energy loss
timescales.  The main problem in applying the method in practice is the
wide range of time-scales, especially for the electron case, ranging
from 10$^4$ years for energy losses to 10$^9$ years for diffusion
around 1 GeV in a large halo.  Use of a time step $\Delta t $
appropropriate to the smallest time-scales guarantees a reliable
solution, but requires a prohibitively large number of steps to reach
the long time-scales.  The following technique was found to work well:
start with a large $\Delta t $ appropriate for the longest scales, and
iterate until a stable solution is obtained. This solution is then
accurate only for cells with $\alpha\ll 1$; for other cells the
solution is stable but inaccurate. Then reduce $\Delta t$ by a factor
(0.5 was adopted) and continue the solution.  This process is repeated
until $\alpha\ll 1$ for all cells, when the solution is accurate
everywhere.  It is found that the inaccurate parts of the solution
quickly decay as soon as the condition $\alpha<1$ is reached for a
cell. As soon as all cells satisfy $\alpha<1$ the solution is continued
with the ADI method to obtain maximum accuracy. A typical run starts
with $\Delta t = 10^9$ years and ends with $\Delta t $ = 10$^4 $
years for nucleons and 10$^2 $ years for electrons performing $\sim$
60 iterations per $\Delta t $.  In this way it is possible to obtain
reliable solutions in a reasonable computer resources, although the CPU
required is still considerable.  All results are output as FITS
datasets for subsequent analysis.

More details, including the software and datasets, can be found on 
the World Wide Web (address available from the authors).

\begin{deluxetable}{llccc}
\tablefontsize{\footnotesize}
\tablecolumns{5}
\footnotesize

\tablecaption{ Coefficients for the Crank-Nicholson method.  \label{table3}}

\tablehead{
\colhead{Process} & \colhead{Coordinate}
 & \colhead{$\alpha_1/\Delta t$}
 & \colhead{$\alpha_2/\Delta t$} & \colhead{$\alpha_3/\Delta t$}
}
\startdata

Diffusion               
   & $ R $         
   & $  \Dxx {2 R_i-\Delta R \over 2 R_i (\Delta R)^2}$ 
   & $  \Dxx {2 R_i\over R_i (\Delta R)^2} $
   & $  \Dxx {2 R_i+\Delta R \over 2 R_i (\Delta R)^2}$   \nl
\noalign{\smallskip}

   & $ z $         
   & $ \Dxx / (\Delta z)^2 $
   & $2\Dxx / (\Delta z)^2 $
   & $ \Dxx / (\Delta z)^2 $  \nl
\noalign{\bigskip}

Convection\tablenotemark{a}
   & $z>0$ ($V>0$)
   & $ V(\zim) / \Delta z $
   & $ V(\zii) / \Delta z $
   & $ 0 $  \nl
\noalign{\smallskip}

   & $z<0$ ($V<0$)
   & $ 0 $
   & $ -{V(\zii) / \Delta z } $
   & $ -{V(\zip) / \Delta z } $  \nl
\noalign{\smallskip}

   & $ p $\ \ ($dV/dz>0$)
   & $ 0 $
   & $ -{1\over 3}\,\pii {dV\over dz} /P^i_{i-1} $
   & $ -{1\over 3}\,\pip {dV\over dz} /P^{i+1}_i $  \nl
\noalign{\bigskip}

\begin{minipage}{2cm} 
Diffusive reacceleration\tablenotemark{a} 
\end{minipage}
   & $ p $
   & $ {2\Dim \over P^{i+1}_{i-1}}({1\over P^i_{i-1}}\! +\! { 2\over\pim})$
   & $ {2 \over P^{i+1}_{i-1}}({\Dip\over P^{i+1}_i}\! +\! 
                               {\Dim\over P^i_{i-1}}) $
   & $ {2\Dip \over P^{i+1}_{i-1}}({1\over P^{i+1}_i}\! -\! { 2\over\pip})$ \nl
\noalign{\bigskip}

Energy loss\tablenotemark{a}
   & $ p $
   & $ 0 $ 
   & $ \dot{p}_{i}/P^{i+1}_i $
   & $ \dot{p}_{i+1}/P^{i+1}_i $\nl
\noalign{\bigskip}

Fragmentation
   & $ R,z,p $
   & $ 0 $
   & $ {1 / 3\tau_f} $
   & $ 0 $ \nl
\noalign{\bigskip}

\begin{minipage}{2cm} 
Radioactive decay
\end{minipage}
   & $ R,z,p $ 
   & $ 0 $ 
   & $ {1 / 3\tau_r} $
   & $ 0 $   \nl

\enddata

\footnotesize 
\tablenotetext{a}{$P^i_j\equiv p_i-p_j$}

\end{deluxetable}

\subsection*{B.1. Diffusion in $R$ \label{diffusion}}
As an example, the coefficients for the radial diffusion term are
derived here.
\begin{equation}
\label{F.1}
{1\over R} {\partial \over \partial R} 
\left(R\, \Dxx {\partial \psi \over \partial R}\right) 
={2 \over R_i} {\Dxx \over R_{i+1}-R_{i-1}} 
\left\{ R_{i+1} {\psi_{i+1}-\psi_i \over R_{i+1}-R_i}
-R_{i-1} {\psi_i-\psi_{i-1} \over R_i-R_{i-1}}\right\}.
\end{equation}
Setting $R_{i+1}-R_i= R_i-R_{i-1}=\Delta R$, one can obtain the
following expressions in terms of our standard form (eq.~[A.1])
\begin{equation}
\label{F.2}
{\alpha_1 \over \Delta t } = \Dxx {2 R_i-\Delta R \over 2 R_i 
   (\Delta R)^2},\quad
{\alpha_2 \over \Delta t}  = \Dxx {2 R_i\over R_i (\Delta R)^2},\quad
{\alpha_3 \over \Delta t}  = \Dxx {2 R_i+\Delta R \over 2 R_i 
   (\Delta R)^2}.
\end{equation}

\pagebreak
\subsection*{B.2. Diffusive reacceleration \label{reacceleration}}
In terms of 3-D momentum phase-space density $f(\vec p)$ the diffusive
reacceleration equation is
\begin{equation}
\label{C.1}
\dfdt= \vec\nabla_p \cdot [ \Dpp\vec\nabla_p\, f(\vec p)] 
= {1\over p^2}\ddp\left[ p^2 \Dpp\dfdp\right] .
\end{equation}
The distribution is assumed isotropic so $f(\vec p) = f(p)$ where
$p=|\vec p|$.  First we rewrite the equation in terms of $\psi(p)= 4\pi
p^2 f(p)$ instead of $f(p)$ and expand the inner differential:
\begin{equation}
\label{C.2}
\dNpdt=\ddp\left[ p^2 \Dpp \ddp{\psi\over p^2 }\right]
=\ddp\, \Dpp \left[ \dNpdp-{2\psi\over p} \right] .
\end{equation}
The differencing scheme is then
\begin{equation}
\label{C.4}
{2\over \pip-\pim}\left[
 \Dip\left({\Nip-\Nii\over\pip-\pii}-{2\Nip\over\pip} \right)  
-\Dim\left({\Nii-\Nim\over\pii-\pim}-{2\Nim\over\pim} \right)
\right] .
\end{equation}
In terms of our standard form (eq.~[A.1]) the coefficients for
reacceleration are
\begin{eqnarray}
\label{C.5}
{\alpha_1 \over \Delta t} &\ =\ & {2\Dim \over \pip-\pim}
\left( {1\over\pii-\pim}+{ 2\over\pim} \right) ,\nonumber\\
{\alpha_2 \over \Delta t} &\ =\ & {2     \over \pip-\pim}
\left( {\Dip \over\pip-\pii}+ {\Dim \over\pii-\pim} \right) ,\\
{\alpha_3 \over \Delta t} &\ =\ & {2\Dip \over \pip-\pim}
\left( {1\over\pip-\pii}-{ 2\over\pip} \right) . \nonumber
\end{eqnarray}

\section*{C. Energy losses}
For nucleon propagation in the ISM the losses are mainly due
to ionization, Coulomb scattering, fragmentation, and radioactive
decay.  For electrons the important processes are ionization, Coulomb
scattering, bremsstrahlung in the neutral and ionized medium, as well
as Compton and synchrotron losses.  Although all these processes are
well-known the formulae for the different cases are rather scattered
throughout the literature and hence for completeness we summarize
the formulae used below.

\begin{figure*}[thb]
\hfill\includegraphics[width=\fwb]{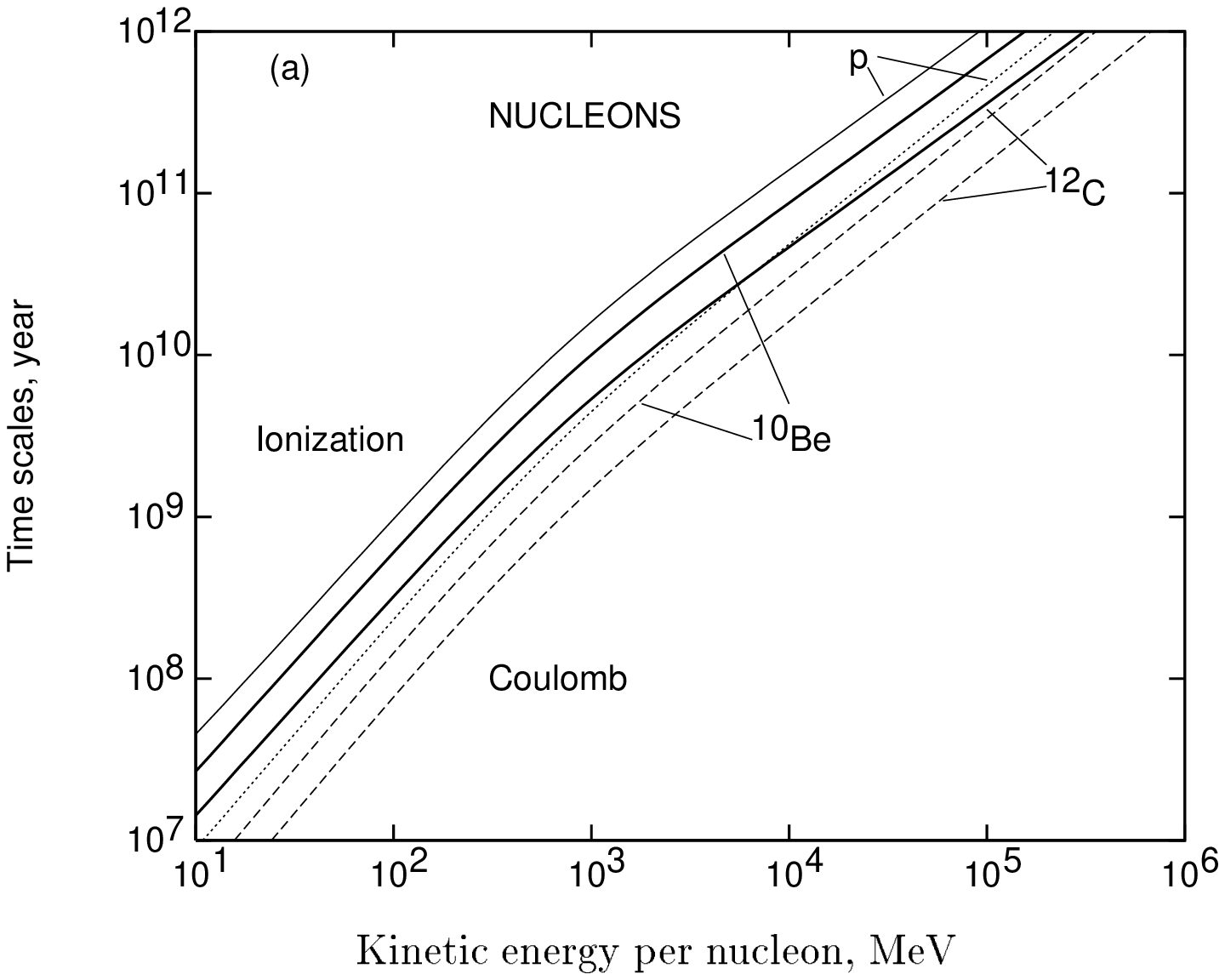}\hfill
      \includegraphics[width=\fwb]{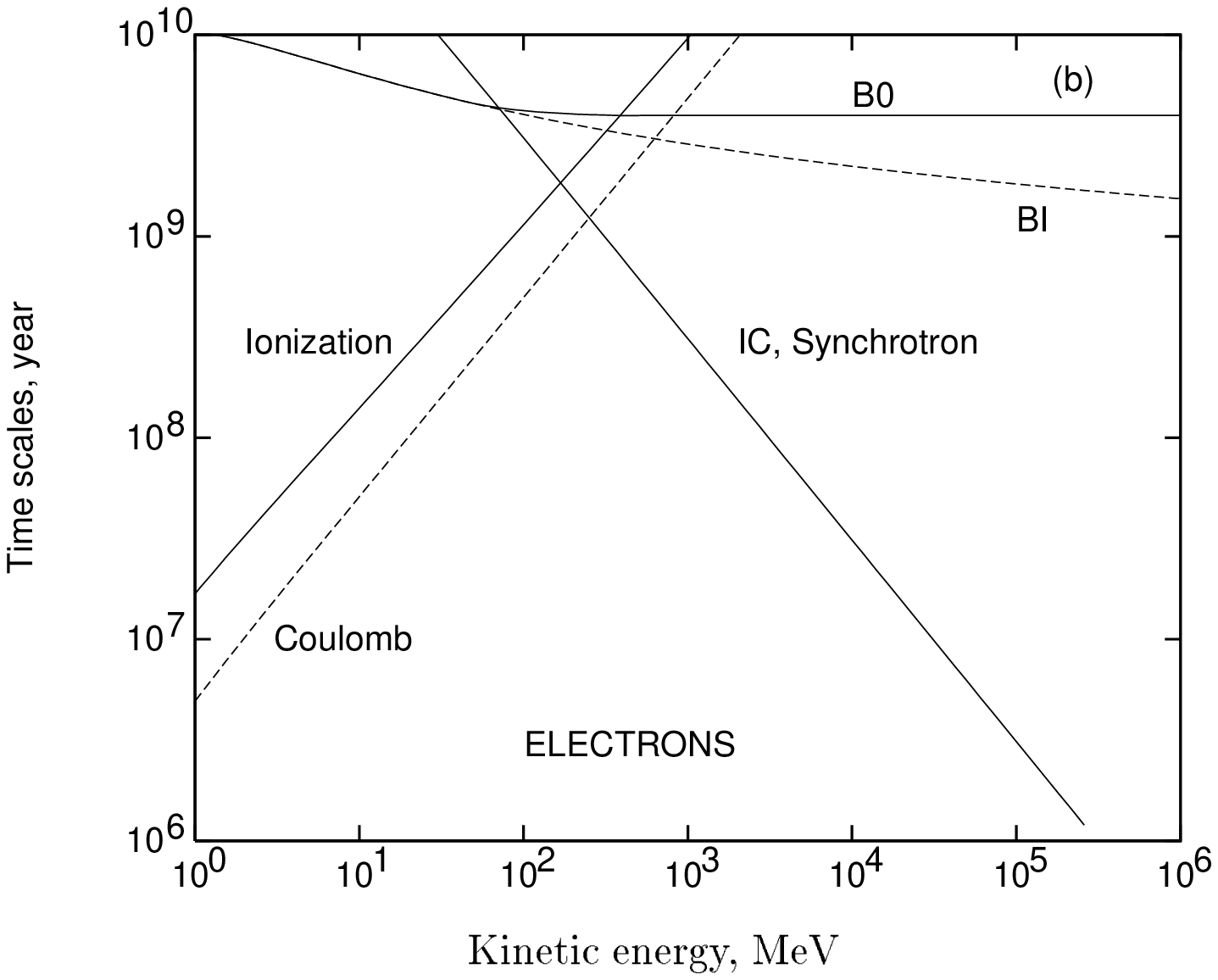}\hfill
\vskip -5pt
\caption[fig13a.ps,fig13b.ps]{ \footnotesize
Energy loss time-scales of (a)
nucleons and (b) electrons in neutral and ionized hydrogen.  The curves
are computed for gas densities $n_H=n_{HII}=0.01$ cm$^{-3}$, and equal
energy densities of photons and magnetic field $U=U_B=1$ eV cm$^{-3}$
(in the Thomson limit).  In panel (a) solid lines correspond to
ionization losses and dashed lines to Coulomb losses (the dotted line
is that for protons).
\label{fig13} }
\end{figure*}

Figure~\ref{fig13} illustrates the energy loss time scales, $E
(dE/dt)^{-1}$, for electrons and nucleons in pure hydrogen. The losses
are shown for equal neutral and ionized gas number densities
$n_H=n_{HII}=0.01$ cm$^{-3}$, and equal energy densities of photons and
the magnetic field $U=U_B=1$ eV cm$^{-3}$ (in the Thomson limit). These
gas and energy densities are chosen to characterize the average values
seen by cosmic-rays during propagation.

\subsection*{C.1. Nucleon energy losses \label{nucleon_energy_losses}}
The Coulomb collisions in a completely ionized plasma are dominated by
scattering off the thermal electrons. The corresponding energy losses
are given by (\cite{MannheimSchlickeiser94}, their eqs.~[4.16],[4.22])
\begin{equation}
\label{D.1}
\left( \frac{dE}{dt}\right) _{\rm Coul}\approx
-4\pi r_e^2 c\, m_ec^2 Z^2 n_e \ln \Lambda\, {\beta^2\over x_m^3+\beta^3},
\end{equation}
where $r_e$ is the classical electron radius, $m_e$ is the electron
rest mass, $c$ is the velocity of light, $Z$ is the {\it projectile}
nucleon charge, $\beta=v/c$ is the nucleon speed, $n_e$ is the electron
number density in plasma, $x_m\equiv (3\sqrt{\pi}/4)^{1/3}\times
\sqrt{2kT_e/m_ec^2}$, and $T_e$ is the electron temperature.
The Coulomb logarithm in the cold plasma limit is given by (e.g.,
\cite{Dermer85})
\begin{equation}
\label{D.5}
\ln\Lambda \approx {1\over 2} \ln 
\left( {m_e^2c^4 \over \pi r_e \hbar^2 c^2 n_e}\cdot
{M \gamma^2\beta^4 \over M+2\gamma m_e} \right) \ ,
\end{equation}
where $\hbar$ is the Planck constant, $M$ is the nucleon mass, and
$\gamma$ is the nucleon Lorentz factor.  For the appropriate number
density, $n_e\sim 10^{-1}-10^{-3}$ cm$^{-3}$, and total energy $E\sim
10^3-10^4$ MeV, the typical value of the Coulomb logarithm $\ln\Lambda$
lies within interval $\sim 40 - 50$, instead of usually adopted value
20.

For the ionization losses we use a general formula
(\cite{MannheimSchlickeiser94}, their eq.~[4.24])
\begin{equation}
\label{D.2}
\left( \frac{dE}{dt}\right) _{\rm I}(\beta\geq\beta_0)=
-2\pi r_e^2 c\, m_ec^2 Z^2 {1\over \beta}
\sum_{s={H,He}} n_s \left[B_s+B'(\alpha_f Z/\beta)\right],
\end{equation}
where $\alpha_f$ is the fine structure constant, $n_s$ is the number
density of the corresponding species in the ISM, $\beta_0=1.4
e^2/\hbar c=0.01$ is the characteristic velocity determined by the
orbital velocity of the electrons in hydrogen, and
\begin{equation}
\label{D.3}
B_s=\left[\ln\left({{2m_ec^2\beta^2 \gamma^2 Q_{\rm max}}\over
{\tilde{I}_s^2}} \right)-2\beta^2-{{2C_s}\over{z_s}} -\delta_s\right],
\end{equation}
where $\gamma$ is the Lorentz factor of the ion.  The largest possible
energy transfer from the incident particle to the atomic electron is
defined by kinematics\footnote{note that there was a typing mistake in
the denominator of the expression given by Mannheim \& Schlickeiser (1994),
which is corrected in our formula.}
\begin{equation}
\label{D.4}
Q_{\rm max}\approx {{2m_ec^2\beta^2\gamma^2}
\over {1+[2\gamma m_e/M]}},
\end{equation}
where $M\gg m_e$ is the nucleon mass, and $\tilde{I}_s$ denotes the
geometric mean of all ionization and excitation potentials of the atom.
Mannheim \& Schlickeiser (1994) give the values $\tilde{I}_{H}=19$ eV
and $\tilde{I}_{He}=44$ eV. The shell-correction term $C_s/z_s$, the
density correction term $\delta_s$, and the $B'$ correction term (for
large $Z$ or small $\beta$) in eqs.~(C.3),(C.4), can be
neglected for our purposes.

Fragmentation and radioactive decay are addressed in
Appendix~A.

\subsection*{C.2. Electron energy losses \label{electron_energy_losses}}
Ionization losses in the neutral hydrogen and helium are given by the
Bethe-Bloch formula (\cite{Ginzburg79}, p.360)
\begin{equation}
\label{E.2}
\left( \frac{dE}{dt}\right) _{\rm I}=-2\pi r_e^2 c\, m_e c^2
{1\over \beta} \sum_{s={H,He}} Z_s n_s
\left[ \ln \left\{ \frac{(\gamma-1) \beta ^2 E^2}
{2I_s^2}\right\} +\frac 18\right] ,
\end{equation}
where $Z_s$ is the nucleus charge, $n_s$ is the gas number density,
$I_s$ is the ionization potential (we use $I_{H}=13.6$ eV, $I_{\rm
He}=24.6$ eV, though the exact numbers are not very important), $E$ is
the total electron energy, $\gamma$ and $\beta=v/c$ are the electron
Lorentz factor and speed, correspondingly.

The Coulomb energy losses in the fully ionized medium, in the cold
plasma limit, are described by (\cite{Ginzburg79}, p.361)%
\begin{equation}
\label{E.3}
\left( \frac{dE}{dt}\right) _{\rm Coul}=-2\pi r_e^2 c\, m_e c^2 Z n\, 
{{1}\over{\beta}} \left[ \ln \left( \frac{E m_ec^2}{4\pi r_e 
\hbar^2 c^2 nZ} \right) -\frac 34\right] ,
\end{equation}
where $Z n\equiv n_e$ is the electron number density.  For an accurate
treatment of the electron energy losses in the plasma of an arbitrary
temperature see, e.g., Dermer \& Liang (1989) and Moskalenko \&
Jourdain (1997).

The energy losses due to $ep$-bremsstrahlung in the cold plasma are
given by the expression (\cite{Stickforth61})
\begin{equation}
\label{E.6}
\left( \frac{dE}{dt}\right) _{ep}=-\frac 23\alpha_f
r_e^2 c\, m_e c^2 Z^2n\left\{ 
\begin{array}{ll}
8\gamma \beta \left[ 1-0.25(\gamma -1)+0.44935(\gamma -1)^2-0.16577(\gamma
-1)^3\right] , & \,\gamma \leq 2; \\ 
\beta ^{-1}\left[ 6\gamma \ln (2\gamma )-2\gamma -0.2900\right] , & 
\,\gamma \geq 2 .
\end{array}
\right.
\end{equation}
For the $ee$-bremsstrahlung one can obtain
(\cite{Haug75,MoskalenkoJourdain97})
\begin{equation}
\label{E.7}
\left( \frac{dE}{dt}\right) _{ee}=-\frac 12\alpha_f r_e^2c\, m_e c^2
Z n\beta \gamma ^{*}Q_{\rm cm}(\gamma ^{*}),
\end{equation}
where 
$$
\begin{array}{l}
\displaystyle{Q_{\rm cm}(\gamma ^{*})=8\frac{p^{*2}}{\gamma ^{*}}\left[
1-\frac{4p^{*}}{ 3\gamma ^{*}}+\frac 23\left( 2+\frac{p^{*2}}{\gamma
^{*2}}\right) \ln (p^{*}+\gamma ^{*})\right]} , \\
\displaystyle{\gamma^{*}=\sqrt{(\gamma +1)/2}}, \\
\displaystyle{p^{*}=\sqrt{(\gamma -1)/2}},
\end{array}
$$
and the asterisk denotes center-of-mass variables. The total
bremsstrahlung losses in the ionized gas is the sum $(dE/dt) _{\rm
BI}=(dE/dt) _{ep}+(dE/dt) _{ee}$.  A good approximation 
gives the expression (\cite{Ginzburg79}, p.408)
\begin{equation}
\label{E.8}
\left( \frac{dE}{dt}\right) _{\rm BI}=-4\alpha_f r_e^2 c\, m_e c^2
Z(Z+1)nE\left[ \ln (2\gamma )-\frac 13\right].
\end{equation}

Bremsstrahlung energy losses in neutral gas can be obtained by
integration over the bremsstrahlung luminosity (\cite{KochMotz59}, see
also \cite{MoskalenkoStrong98b})
\begin{equation}
\label{E.4}
\left( \frac{dE}{dt}\right) _{\rm B0}= -c \beta \sum_{s={H,He}} n_s
\int dk\, k \frac{d\sigma_s }{dk }.
\end{equation}
A suitable approximation (max 10\% error near $E \sim 70$ MeV) for
eq.~(C.11) gives the combination (cf.\ eq.~[C.10])
\begin{equation}
\label{E.5}
\left( \frac{dE}{dt}\right) _{\rm B0}=
\left\{
\begin{array}{ll}
\displaystyle -4\alpha_f r_e^2 c\, m_e c^2 E
\left[\ln (2\gamma )-\frac 13\right]\sum_{s={H,He}}n_s Z_s(Z_s+1) ,
& \ \gamma\la 100; \\
\displaystyle -c\, E \sum_{s={H,He}} {\frac{n_s M_s}{T_s}},
& \ \gamma\ga 800,
\end{array}
\right.
\end{equation}
(see \cite{Ginzburg79}, p.386, 409), with a linear connection in
between. Here $M_s$ is the atomic mass, and $T_s$ is the radiation
length ($T_H\simeq 62.8$ g/cm$^2$, $T_{He}\simeq 93.1$ g/cm$^2$).

The Compton energy losses are calculated using the Klein-Nishina cross
section (\cite{Jones65,MoskalenkoJourdain97})
\begin{equation}
\label{E.11}
\frac{dE}{dt}=\frac{\pi r_e^2 m_ec^2 c}{2\gamma^2\beta}\int_0^\infty 
d\omega\,f_\gamma(\omega)[S(\gamma,\omega,k^+)-S(\gamma,\omega,k^-)],
\end{equation}
where the background photon distribution, $f_\gamma(\omega)$, is
normalized on the photon number density as $n_\gamma=\int
d\omega\,\omega^2f_\gamma(\omega)$, $\omega$ is the energy of the
background photon taken in the electron-rest-mass units,
$k^\pm=\omega\gamma(1\pm\beta)$,
\begin{eqnarray}
\label{E.12}
S(\gamma,\omega,k)&=&\omega
\left\{ \left(k+\frac{31}{6}+\frac{5}{k}+\frac{3}{2k^2}\right)\ln(2k+1)
-\frac{11}{6}k-\frac{3}{k}+\frac{1}{12(2k+1)}+\frac{1}{12(2k+1)^2}
+Li_2(-2k)
\right\} \nonumber \\
&&-\gamma
\left\{ \left(k+6+\frac{3}{k}\right)\ln(2k+1)
-\frac{11}{6}k+\frac{1}{4(2k+1)}-\frac{1}{12(2k+1)^2}
+2Li_2(-2k)
\right\}, 
\end{eqnarray}
and $Li_2$ is the dilogarithm
\begin{eqnarray}
\label{E.13}
Li_2(-2k)&=&-\int_0^{-2k}dx\,\frac{1}{x}\ln(1-x) \\
         &=&\left\{
\begin{array}{ll}
\sum_{i=1}^\infty(-2k)^i/i^2, & k\le0.2; \\
-1.6449341+{1\over2}\ln^2 (2k+1)-\ln (2k+1)\ln (2k)+\sum_{i=1}^\infty
i^{-2}(2k+1)^{-i}, & k\ge0.2.
\end{array} \right. \nonumber
\end{eqnarray}

The synchrotron energy losses are given by
\begin{eqnarray}
\label{E.10}
\left( \frac{dE}{dt}\right) _{\rm S}&=&-\frac{32}9\pi r_e^2 c\,U_B
\gamma ^2 \beta ^2, 
\end{eqnarray}
where $U_B=\frac{H^2}{8\pi}$ is the energy density of the random
magnetic field.

\twocolumn


\end{document}